\documentclass[a4paper,cleveref,USenglish]{lipics-v2021}

\pdfoutput=1

\usepackage{algorithm}
\usepackage[noend]{algpseudocode}  % part of algorithmicx package

\usepackage{eqparbox} % for aligned comments
\usepackage{amsthm}
\usepackage{graphicx}
\usepackage{xcolor}
\usepackage{enumitem}
\usepackage{subcaption} % for captioning parts of figures
\usepackage{soul} % for highlighting text
\usepackage{combelow} % for comma below s in name Patrascu
\usepackage{cite} % for multiple citation with different postnotes for each using \cites

\hideLIPIcs  %uncomment to remove references to LIPIcs series (logo, DOI, ...), e.g. when preparing a pre-final version to be uploaded to arXiv or another public repository
\nolinenumbers

\crefalias{part}{enumi}
\crefname{part}{part}{parts}
\Crefname{part}{Part}{Parts}

\newtheorem{invariant}[theorem]{Invariant}

\crefname{invariant}{invariant}{invariants}
\Crefname{invariant}{Invariant}{Invariants}
\usepackage{etoolbox}

\newcommand{\algstrut}[1][\algruledefaultfactor]{\vrule width 0pt
depth .25\baselineskip height #1\baselineskip\relax}
\newcommand*{\algrule}[1][\algorithmicindent]{\hspace*{.5em}\vrule\algstrut
\hspace*{\dimexpr#1-.5em}}

\makeatletter
\newcount\ALG@printindent@tempcnta
\def\ALG@printindent{%
    \ifnum \theALG@nested>0% is there anything to print
    \ifx\ALG@text\ALG@x@notext% is this an end group without any text?
    \else
    \unskip
    \ALG@printindent@tempcnta=1
    \loop
    \algrule[\csname ALG@ind@\the\ALG@printindent@tempcnta\endcsname]%
    \advance \ALG@printindent@tempcnta 1
    \ifnum \ALG@printindent@tempcnta<\numexpr\theALG@nested+1\relax% can't do <=, so add one to RHS and use < instead
    \repeat
    \fi
    \fi
}%

\patchcmd{\ALG@doentity}{\noindent\hskip\ALG@tlm}{\ALG@printindent}{}{\errmessage{failed to patch}}

\AtBeginEnvironment{algorithmic}{\lineskip0pt}

\algnewcommand\StartFromLine[1]{\setcounter{ALG@line}{\numexpr#1-1}}

\ifthenelse{\equal{\ALG@noend}{t}}{%
	\algdef{SE}[FUNCTION]{Function}{EndFunction}[3]{\op{#2}\ifthenelse{\equal{#3}{}}{}{(#3)}\ifthenelse{\equal{#1}{}}{}{ : #1}}{}%
	\algtext*{EndFunction}%
	}{% if using ends, you have to add function name in braces after \EndFunction
	\algdef{SE}[FUNCTION]{Function}{EndFunction}[3]{{\op{#2}}\ifthenelse{\equal{#3}{}}{}{(#3)}\ifthenelse{\equal{#1}{}}{}{ : #1}}[1]{\algorithmicend\ \op{#1}}%
	}
	
\algblockdefx{Rpt}{EndRpt}[1]{\algorithmicrepeat\ #1}{\algorithmicend\ \algorithmicrepeat}

\newcommand{\algorithmictype}{\textbf{type}}
\ifthenelse{\equal{\ALG@noend}{t}}{%
	\algdef{SE}[TYPE]{Type}{EndType}[1]{\algorithmictype\ #1}{}%
	\algtext*{EndType}%
	}{% if using ends, you have to add type name in braces after \EndType
	\algdef{SE}[TYPE]{Type}{EndType}[1]{\algorithmic type\ #1}[1]{\algorithmicend\ \op{#1}}%
	}

\makeatletter%these three lines resize code to small font
\renewcommand{\ALG@beginalgorithmic}{\small}
\makeatother

\definecolor{lightgrey}{gray}{0.87}
\DeclareRobustCommand{\hlgrey}[1]{{\sethlcolor{lightgrey}\hl{#1}}}
\newcommand{\newcode}[1]{\hlgrey{#1}}

\makeatletter
\patchcmd{\ALG@doentity}{\item[]\nointerlineskip}{}{}{}
\makeatother

\makeatletter
\newcommand{\citecomment}[2][]{\citen{#2}#1\citevar}
\newcommand{\citeone}[1]{\citecomment{#1}}
\newcommand{\citetwo}[2][]{\citecomment[,~#1]{#2}}
\newcommand{\citevar}{\@ifnextchar\bgroup{;~\citeone}{\@ifnextchar[{;~\citetwo}{]}}}
\newcommand{\citefirst}{\@ifnextchar\bgroup{\citeone}{\@ifnextchar[{\citetwo}{]}}}
\newcommand{\cites}{[\citefirst}
\makeatother

\newcommand{\op}[1]{\mbox{\textsf{#1}}}
\newcommand{\boldop}[1]{\mbox{\textbf{\textsf{#1}}}}
\newcommand{\CAS}{\op{CAS}}

\newcommand{\x}[1]{\mbox{\textit{#1}}} % for multi-chararacter identifiers (variable names, etc.)

\newcommand{\floor}[1]{\lfloor #1 \rfloor}

\newcommand{\nil}{\op{Nil}}

\newcommand{\true}{\mbox{\textsf{true}}}
\newcommand{\false}{\mbox{\textsf{false}}}

\newcommand{\wasOnPath}{Lemma 20} % Lemma \ref{was-on-search-path}
\newcommand{\realTime}{Lemma 27} % Lemma \ref{real-time}
\newcommand{\bstProperty}{Lemma 21} % invariant about BST property
\newcommand{\stillonpath}{Lemma 19} % Lemma \ref{still-on-path}
\newcommand{\changeUnflagged}{Lemma 14 and 17} 

\title{Lock-Free Augmented Trees}
\author{Panagiota Fatourou}{FORTH ICS an University of Crete, Greece}{faturu@ics.forth.gr}{https://orcid.org/0000-0002-6265-6895}{}
\author{Eric Ruppert}{York University, Canada}{ruppert@eecs.yorku.ca}{https://orcid.org/0000-0001-5613-8701}{}
\authorrunning{P.\ Fatourou and E.\ Ruppert}
\begin{document}

\maketitle

% !TEX root =  augmented-tree.tex

\abstract{%
Augmenting an existing sequential data structure with extra information to support
greater functionality is a widely used technique.
For example, search trees are augmented to build sequential data structures like order-statistic trees,
interval trees, tango trees, link/cut trees and many others.

%A classical data structure technique is to augment
%the nodes of a search tree with extra fields
%to support additional operations.
%This technique is used to build sequential data structures like order-statistic trees,
%interval trees, tango trees, link/cut trees and many others.

We study how to design \emph{concurrent} augmented tree data structures.
We present a new, general technique that can augment a lock-free tree to add any new fields to each tree node, provided the new fields' values
can be computed from information in the node and its children.
This enables the design of lock-free, linearizable analogues of a wide variety of classical augmented data structures.
As a first example, we give a wait-free trie that stores a set $S$ of elements drawn from $\{1,\ldots,N\}$
and supports linearizable order-statistic queries such as finding the $k$th smallest element of $S$.
Updates and queries take $O(\log N)$ steps.
We also apply our technique to a lock-free binary search tree (BST), where changes to the structure of the tree make the linearization argument more challenging.  Our augmented BST supports order statistic 
queries in $O(h)$ steps on a tree of height $h$.
The augmentation does not affect the asymptotic running time of the updates.

For both our trie and BST, we give an alternative augmentation to improve searches and order-statistic queries to run in $O(\log |S|)$ steps (with a small increase in step complexity of updates).
As an added bonus, our technique supports arbitrary multi-point queries (such as range queries)
with the same time complexity as they would have in the corresponding sequential data structure.

}

% !TEX root =  augmented-tree.tex

\section{Introduction}
\label{introduction}

Augmentation is a fundamental technique to enhance the functionality of sequential data structures
and to make them more efficient, particularly for queries.
Augmentation is sufficiently important to warrant a chapter in the algorithms
textbook of Cormen et al.~\cite{CLRS4-17}, which illustrates the technique
with the most well-known example
of augmenting a binary search tree (BST) to support many additional operations.
By \emph{augmenting} each node to store the size of the subtree rooted at that node,
the BST can answer order-statistic queries, such as finding the $j$-th smallest element in the BST or
the rank of a given element, %the maximum, minimum or median element, 
%$j$-th successor of an element 
\emph{in sub-linear time}.
In a balanced BST, these queries take logarithmic time
whereas a traversal of an unaugmented BST would take linear time to answer them.

More generally,  
each node of a (sequential) tree data structure can be augmented with
any number of additional fields that are useful for various applications,
provided that the new fields of a node can be computed using information in that node and its children.
When applied to many standard trees, such as balanced or unbalanced BSTs, tries or B-trees,
the augmentation does not affect the asymptotic time for simple updates, like insertions or deletions,
but it can facilitate many other efficient operations.
For example, a balanced BST of keys can be augmented for a \op{RangeSum} query that computes
the sum of all keys within a given range in logarithmic time 
by adding a field to each node that stores the sum of keys in the node's subtree.  
(The sum can be replaced by any associative aggregation operator, such as minimum, maximum or product.)
Similarly, a BST of key-value pairs can be augmented to aggregate the \emph{values} associated with keys 
in a given range:  each node should store the sum of values in its subtree.
One can also filter values, for example to obtain the aggregate of all odd values within a range.
More sophisticated augmentations can also be used.
For instance, an interval tree stores a set of intervals in a balanced BST sorted by the left endpoints, where each node
is augmented to store the maximum right endpoint of any interval in the node's subtree,
so that one can determine whether any interval in the BST includes a given point in logarithmic time\cite{CLRS4-17}.
There are many other types of augmented trees, including one representing piecewise constant functions \cites{BKM+03}[Section 4.5]{Bra08}, measure trees \cite{GMW83}, priority search trees \cite{McC85} and segment trees~\cite{Bendl77,Berg08}.
Section \ref{faster-queries} gives another novel example of how to use tree augmentation.
Augmented trees are also used as a building block for many other sequential data structures
such as %multidimensional range trees \here{\cite{?}},
link/cut trees \cite{ST83} and tango trees \cite{DHIP07}. %\here{Maybe get a better list of applications for camera-ready}
These structures have many applications in graph algorithms, computational geometry and
databases.
%I replaced the following by the more general sentence above, because I thought it was repetitive
% to mention multidimensional range trees and interval trees, and then say they could be used for k-dimensional
% range queries and interval queries.
%
%Through augmentation, we can also efficiently solve a wide collection of problems,
%such as answering interval queries, k-dimensional range queries, or windowing queries, 
%designing weighted inverted indices, solving the segment or the rectangle intersection problem, 
%or the point location  problem, determining range overlaps, and many others. 

We consider how to augment \emph{concurrent} tree data structures.
The resulting data structures are linearizable and lock-free and use single-word compare-and-swap (\op{CAS}) instructions.
The technique we introduce is very general:  as in the sequential setting, it can handle
any augmentation to a lock-free tree data structure where the new fields can be computed using the data stored in the node
and its children. 
Thus, it can be used to provide efficient, lock-free shared implementations of many of the
sequential data structures mentioned above.
Our augmentation does not affect the asymptotic running time of update operations.
Moreover, we provide a way for queries to obtain a snapshot of the data structure
so that they can simply execute the sequential code to answer the query.

\begin{figure}
\begin{subfigure}[b]{0.21\textwidth}
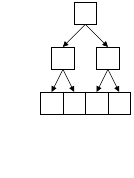
\caption{Non-concurrent static trie for the set $S=\{1,2,3\}$.}\label{bitvector}
\end{subfigure}
\hfill
\begin{subfigure}[b]{0.38\textwidth}
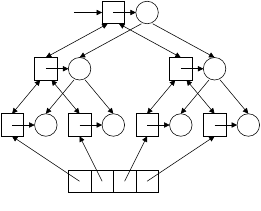
\caption{Initialization of concurrent trie.}\label{init-fig}
\end{subfigure}
\hfill
\begin{subfigure}[b]{0.38\textwidth}
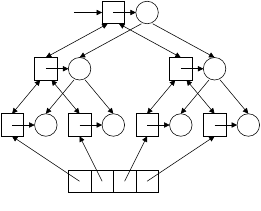
\caption{Concurrent trie for set $S=\{1, 2,3\}$.}\label{fig123}
\end{subfigure}
\caption{Examples of the trie data structure when $U=\{0,1,2,3\}$.
Nodes are shown as squares, Versions are shown as circles containing their 
$sum$ fields.\label{tree-examples}}
\end{figure}

For ease of presentation, we first illustrate the technique 
on top of a simple data structure that represents a dynamically-changing set $S$ of keys 
drawn from the universe $U=\{0,\ldots,N-1\}$.
The basic data structure is a {\em static} binary trie of height $\log_2 N$, where each key of $U$ is assigned a leaf.
Our technique mirrors this tree of nodes by a tree of Version objects, which store the mutable fields 
of the augmentation.
See \Cref{fig123} for an example where each node stores the number of elements present in its subtree to support
order-statistic queries.
Insertions and deletions of elements modify the appropriate leaf of the tree (and its Version), and then
cooperatively propagate any changes to the Version objects stored in ancestors of that leaf until reaching the root.
This cooperative approach ensures updates perform a constant number of steps at each node along this path, so they take $O(\log N)$ steps in total.
Moreover, we design the updates so that the fields of Version objects (including their child pointers) never change.
Thus, reading the root node's Version object provides a ``snapshot'' of the entire Version tree, which a query can
then explore at its leisure, knowing that it will not be changed by any concurrent update operations.
This allows any query operation that follows pointers from the root to be performed exactly as  in a 
sequential version of the  data structure, using the same number of steps.
For example, order-statistic queries can be answered using $O(\log N)$ steps, and the size of $S$ 
can be found in $O(1)$ steps.  All operations are wait-free.
%\here{Can it be that the reference to versions here is confusing. We actually do not use the multi-versioning technique. 
%I think it would be nice if this becomes clear early in the paper.}
%\here{I disagree: we do use a kind of multiversioning.  Here is the definition of multiversion concurrency control from wikipedia, which sounds exactly like what we do:
%"When an MVCC database needs to update a piece of data, it will not overwrite the original data item with new data, but instead creates a newer version of the data item. Thus there are multiple versions stored. The version that each transaction sees depends on the isolation level implemented. The most common isolation level implemented with MVCC is snapshot isolation. With snapshot isolation, a transaction observes a state of the data as of when the transaction started."}

Then, we describe how to apply the technique to a BST.
This encompasses additional complications because the structure of the tree
changes dynamically as keys are inserted or deleted.
We base our lock-free augmented BST on the lock-free BST of Ellen et al.~\cite{EFHR14},
whose amortized step complexity is $O(h+c)$ per operation, where $h$ is  the height of the tree
and $c$ is the point contention (i.e., the maximum number of update operations running concurrently).
Our augmentation does not affect this asymptotic step complexity of the lock-free update operations,
and wait-free queries can again be performed using the same number of steps as in a sequential implementation.
%\here{mention that it should work with other trees like Mittal/Natarajan, trie}

In an augmented tree, each insertion or deletion must typically modify many tree nodes.  For example,
an insertion in an order-statistic tree must increment the count in all ancestors of the  inserted node.
In the concurrent setting, we must ensure that all of these changes 
appear to take place atomically, so that queries operate correctly.
It is generally very difficult to design lock-free\linebreak
data structures
where many modifications must appear atomic.  Our proposed technique
addresses this challenge in a rather simple way.  However, the full
proofs of correctness are fairly challenging.

Whether augmentation of the tree is needed or not, our technique also
provides a simple way of taking a snapshot of the tree  to answer queries
that must examine multiple locations in the tree, such as a range query.
Thus, in addition to supporting augmentation, our technique provides an alternative approach to other recent work on providing linearizable range queries on concurrent trees \cite{AB18,BBB+20,BA12,Cha17,FPR19} 
or more general snapshots \cite{NHP22,PBBO12,WBB+21}.
Ordinarily, our snapshots can be discarded when the query completes, but if desired,
they can also be used to maintain past versions of the data structure.
Many of these other approaches use multiversioning in a way that requires complex
schemes for unlinking old, obsolete versions from the data structure to facilitate garbage collection
(e.g., \cite{BBF+21,WBFR23}).
The simplicity of our approach avoids this.

Similarly, our approach yields a constant-time algorithm for finding the number of keys in a lock-free tree.
In contrast, a previous, more general method for adding a size query 
to any dynamic set \cite{SP22} is substantially more complicated, and size queries take $\Omega(P)$ steps
in a system of $P$ processes.

\section{Related Work}
\label{related}

There is very little previous work on lock-free implementations of augmented trees.
In a recent manuscript, Kokorin, Alistarh and Aksenov \cite{KAA23}
describe a wait-free BST supporting order-statistic queries and range queries.
They use a FIFO queue for each tree node.
Before reading or writing a node, an operation must join the node's queue
and help each operation ahead of it in the queue
by performing that operation's access to the node and, if necessary, adding the operation to the queue of the node's child (or children).
This adds $\Omega(Ph)$ to the worst-case step complexity per operation when there
are $P$ processes accessing a tree of height~$h$.
To handle order-statistic queries, each node stores the size of its subtree.
They assume that all updates will succeed (e.g., a \op{Delete}($k$) will always 
find $k$ to be in the BST), so that an update can modify the \x{size} field of nodes on its 
way down a path in the tree.
%It is not clear in \cite{KAA23}  how a helper of a \op{Delete}($k$) at the root
%can know whether to decrement the root's size without first completing the entire deletion to see 
%whether $k$ was present.  Moreover,
This top-down approach does not seem to generalize 
to other augmentations where new fields are generally computed bottom-up because
the values of the fields of a node usually depend on the values in the node's children.

Sun, Ferizovic and Blelloch \cite{SFB18}  discuss augmented trees in a parallel setting,
but their focus~is on processes sharing the work of a single expensive operation
(like a large range query or unioning two trees), whereas our goal is to support
multiple concurrent operations on the data structure.

% PO15 and Pro15 are not relevant at all

The cooperative approach we use to propagate operations up to the root of the tree
originates in the universal construction of Afek, Dauber and Touitou \cite{ADT95}.
It has been used to build a variety of lock-free data structures \cite{AR23,FK20,JP05,NR23}.
All of these applied the technique to a tournament tree with one leaf per process.
A process adds an operation at its leaf, and processes move up the tree gathering larger  batches of operations
until the batch is applied to the data structure at the root of the tournament tree. 
Here, we instead apply the approach directly to the tree data structure itself to build larger and larger pieces of 
the updated tree until we reach the root, at which time we have constructed a new version of the data structure
(without destroying any previous versions).

Jayanti \cite{Jay02} used the technique of \cite{ADT95}
to implement an array $A[1..n]$ where processes
can update an array element and query the value of some fixed function
$f(A[1],\ldots,A[n])$, if $f$ can be represented as an evaluation tree similar to a circuit
(where leaves are elements of the array, each internal node represents some function of its children
and the root represents $f$).
Updates cooperatively propagate changes up the tree so that a query can read $f$'s value 
from the root.
Our trie has some similarities, but is much
more general: instead of simply computing a
function value, we construct a copy of the data structure that can be used for more complex
queries.  Our BST implementation goes further to remove the restriction that the
shape of the tree being used for the propagation is fixed.

A different technique that also involves cooperatively building trees bottom-up also appears in Chandra, Jayanti and Tan's
construction \cite{CJT98} of closed objects (where the effect of any pair of operations is equivalent to another operation).
They build trees that represent batches of operations to keep track of the sequence of all operations applied
to the closed object being implemented.
In contrast, we  directly build a representation of the implemented tree data structure.

Our work is on augmenting tree data structures with additional
fields to support additional functionality.
The main challenge is to make changes to several nodes required by an
insert or delete appear atomic.
As a byproduct, our technique for doing this also allows processes
to take a snapshot of the tree, which can be used to answer arbitrary
queries on the state of the tree.  For example, it can be used on a BST to find all keys in a given range.
A number of recent papers \cite{BW24,FPR19,NHP22,PBBO12,WBB+21} use some form of multiversioning to
add the ability to take a snapshot of the state of a concurrent data structure
(but without addressing the problem of augmentation).
Our approach applies only to trees, whereas some of the other work can be applied to arbitrary data structures,
but we do get more efficient queries:  a query in our scheme has the same 
step complexity as the corresponding
query in a sequential implementation, whereas a query 
that runs on top of other multi-versioning schemes, such as that of \cite{WBB+21},
can take additional steps for
every update to the tree that is concurrent with the query.
Our approach is more akin to that of functional updates to the data structure that leave old versions accessible, 
as in the work on classical persistent data structures \cite{DSST89}, but the novelty
here is that the new versions are built cooperatively by many concurrent operations.

% !TEX root =  augmented-tree.tex

\section{Augmented Static Trie}
\label{static}

In this section, we illustrate our augmentation technique for the simple data structure that represents a set $S$ of keys drawn from the universe $U=\{0,1,\ldots,N-1\}$.
%We first illustrate our augmentation technique for a simple data structure that represents a set $S$ of keys drawn from the universe $U=\{0,1,\ldots,N-1\}$.
For simplicity, assume $N$ is a power of 2.
A simple, classical data structure for $S$ is a 
bit vector $B[0..N-1]$, where $B[i]=1$ if and only if $i\in S$.
Even in a concurrent setting, update operations (insertions and deletions of keys) 
can be accomplished by a single \CAS\ instruction
and searches for a key by a single read instruction.

%Now, suppose we wish to add support for the following order-statistic queries.
To illustrate the technique, suppose we wish to support the following order-statistic queries.
\begin{itemize}
\item
\op{Select}$(k)$ returns the $k$th smallest element in $S$.
\item
\op{Rank}$(x)$ returns the number of elements in $S$ smaller than or equal to $x$.
\item
\op{Predecessor}$(x)$ returns the largest element in $S$ that is smaller than $x$.
\item
\op{Successor}$(x)$ returns the smallest element in $S$ that is larger than $x$.
\item
\op{Minimum} and \op{Maximum} return the smallest and largest element in $S$.
\item
\op{RangeCount}($x_1,x_2$) returns the number of elements  in $S$ between $x_1$ and $x_2$.
\item
\op{Size} returns $|S|$, the number of elements in the set $S$.
\end{itemize}
In the \emph{non-concurrent} setting we can build a binary tree of height $\log_2 N$ whose leaves 
correspond to the elements of the bit vector, as shown in \Cref{bitvector}.
We augment each node $x$ with a \x{sum} field to store the sum of the bits in $x$'s descendant leaves, i.e., the number of elements of $S$ in the subtree rooted at $x$.
For a leaf, the \x{sum} field is simply the bit that indicates
if that leaf's key is present in $S$.
The \x{sum} field of an internal node can be computed as the sum of its children's \x{sum} fields.  
It is straightforward to see that \op{Size} queries can then be answered in $O(1)$ time and 
the other order-statistic queries can be answered in $O(\log N)$ time.
We call this data structure a \emph{static trie} because the path to the leaf for $i\in U$
is dictated by the bits of the binary representation of $i$, as in a binary trie \cites{Fre60}[Section 6.3]{TAOCP-3}:  starting from the root, go left
when the next bit is 0, or right when the next bit is 1.  Although the trie's \emph{shape} is static,
it represents a dynamically changing set~$S$.

\subsection{Wait-Free Implementation}
\label{static-alg}

The challenge of making the augmented trie concurrent is that each 
insertion or deletion, after setting the bit in the appropriate
leaf, must update the \x{sum} fields of all ancestors of that leaf.
All of these updates must appear to take place atomically.
To achieve this, we use a  modular design that separates the structure of the tree (which is immutable) from
the mutable \x{sum} fields of the nodes.  This modularity means the same approach can be used
to augment various kinds of lock-free trees.

We use Node objects to represent the tree structure.  \x{Root} is a shared pointer to the root Node. 
To expedite access to the leaves, we use an array $\x{Leaf}[0..N-1]$,
where $\x{Leaf}[i]$ stores a pointer to the leaf Node for key $i$.
Each Node has a  \x{version} field,
which stores a pointer to a Version object that contains the current value of the Node's \x{sum} field.
A Version object $v$ associated with a node $x$ also stores pointers \x{left} and \x{right} to
the Version objects that were associated with $x$'s children at the time when $v$ was
created.
This way, the Version objects form a \emph{Version tree} whose shape mirrors the
tree of Nodes.  See \Cref{init-fig}. 
Query operations are carried out entirely within this Version tree. 
To simplify queries,  fields of Versions are immutable,
so that when a query reads \x{Root}.\x{version},
it obtains a snapshot of the entire Version tree that 
it can later explore by following child pointers.

To see how updates work, consider an \op{Insert}($3$) operation,
starting from the initial state of the trie shown in \Cref{init-fig}.
It must increment the \x{sum} field of the leaf for key $i$ and of each Node along the path
from that leaf to the root.
Since Versions' fields are immutable,
whenever we wish to change the data in the Version associated with a Node $x$,
we create a \emph{new} Version initialized with the new \x{sum} value for the Node, together with the pointers to the two Versions of  $x$'s children from which $x$'s \x{sum} field was computed.
Then, we use a \CAS\ to attempt to swing the pointer in $x.\x{version}$
to the new Version.
If the \op{Insert}($3$) runs by itself, it would make the sequence of changes shown in \Cref{insert3-fig} as it works its way up the tree.
The \op{Insert} is linearized when \x{Root}.\x{version} is changed (\Cref{insert3-c}).
Prior to that linearization point, any query operation reading the root's \x{version} field
gets a pointer to the root of the initial Version tree; after it, a query operation gets 
a pointer to a Version tree that reflects all the changes required by the \op{Insert}.
A \op{Delete}($k$) operation is handled similarly by decrementing the \x{sum} field at each Node along 
the path from $k$'s leaf Node to the root.

Now, consider concurrent updates.
Each update operation must ensure that the root's \x{version} pointer is updated to reflect the effect of the update.
We avoid the performance bottleneck that this could create
by having update operations \emph{cooperatively} update Versions.
At each Node $x$ along the leaf-to-root path,
the update reads the \x{version} field from both of $x$'s children, 
creates a new Version for $x$ based
on the information in the children's Versions,
and attempts to install a pointer to it in $x.\x{version}$ using a \CAS.
Following the terminology of \cite{Jay02}, we call this procedure a \emph{refresh}.%
This approach is cooperative, since a refresh of Node $x$ by one update will propagate
information from all updates that have reached either child of $x$ to $x$.
If an update's first refresh on $x$ fails, it performs a second refresh.
This is called a \emph{double refresh} of $x$.
We shall show that attempting a refresh twice at each Node suffices:  if 
both of the \op{CAS} steps in an update's \emph{double refresh} on a Node $x$ fail,
it is guaranteed that some other process has propagated the update's information to $x$.
%The double refresh mechanism originally appeared in Afek, Dauber and Touitou's universal construction~\cite{ADT95},
%although they applied it to a tournament tree representing processes, rather than to a 
%data structure directly.\here{The last sentence can be removed. It already appears in Related Work.}

\begin{figure}[t]
\begin{algorithmic}[1]
\Type Node \Comment{used to store nodes of static trie structure}
	\State Node* \x{left}, \x{right} \Comment{immutable pointers to children Nodes}
	\State Node* \x{parent}		\Comment{immutable pointer to parent Node}
	\State Version* \x{version}  \Comment{mutable pointer to current Version}	
\EndType

\medskip

\Type Version \Comment{used to store a Node's augmented data}
	\State Version* \x{left}, \x{right} \Comment{immutable pointers to children Versions}
	\State int \x{sum} \Comment{immutable sum of descendant leaves' bits}
\EndType
\end{algorithmic}
\caption{Object types used in wait-free trie data structure.\label{trie-types}}
\end{figure}

\Cref{trie-types} describes the fields of our objects. \Cref{pseudocode} provides pseudocode for the implementation.
It is substantially simpler than previous lock-free tree data structures for sets, even though
it includes augmentation and provides atomic snapshots.
In our code, if \x{ptr} is a pointer to an object $O$, \x{ptr}.\x{f} denotes field $f$ of $O$.
%The initialization of our data structure as shown in \Cref{init-fig} is described at lines \ref{init-code}--\ref{init-code-end}.
%\here{I would also remove this sentence to save space.}
A shared pointer \x{Root} points to the root Node of the binary tree with $N$ leaves.
We use an array $\x{Leaf}[0..N-1]$ where $\x{Leaf}[k]$ points to the leaf Node
for key $k$.

\begin{figure}
\begin{algorithmic}[1]
\StartFromLine{8}
\State Initialization (refer to Figure \ref{init-fig}):\label{init-code}
\State Node* $\x{Root}\leftarrow$ root of a perfect binary tree of Nodes with $N$ leaves.
\State For each Node $x$, $x.\x{version}$ points to a new Version with fields $\x{sum} \leftarrow 0$, $\x{left} \leftarrow x.\x{left}.\x{version}$\\
\hspace*{25mm} and $\x{right} \leftarrow x.\x{right}.\x{version}$.
\State Node* $\x{Leaf}[0..N-1]$ contains pointers to the leaf Nodes of the binary tree.\label{init-code-end}

\medskip

\Function{Boolean}{Insert}{int $k$} \Comment{Add $k$ to $S$; return \true\ iff $k$ was not already in $S$}
	\State $\x{old} \leftarrow \x{Leaf}[k].\x{version}$ \label{read-old-ins}
	\State $\x{result} \leftarrow (\x{old}.\x{sum} = 0)$\label{test-ins}
	\If{\x{result}}
		\State $\x{new} \leftarrow$ new Version with $\x{sum}\leftarrow 1$, $\x{left} \leftarrow\nil$, and $\x{right} \leftarrow \nil$\label{new-leaf-ins}
		\State $\x{result} \leftarrow \CAS(\x{Leaf}[k].\x{version}, \x{old}, \x{new})$\label{cas-ins}
	\EndIf
	\State $\op{Propagate}(\x{Leaf}[k].\x{parent})$\label{prop-ins}
	\State\Return \x{result}
\EndFunction

\medskip

\Function{Boolean}{Delete}{int $k$} \Comment{Remove $k$ from $S$; return \true\ iff $k$ was in $S$}
	\State $\x{old} \leftarrow \x{Leaf}[k].\x{version}$ \label{read-old-del}
	\State $\x{result} \leftarrow (\x{old}.\x{sum} = 1)$\label{test-del}
	\If{\x{result}}
		\State $\x{new} \leftarrow$ new Version with $\x{sum}\leftarrow 0$, $\x{left}\leftarrow\nil$ and $\x{right}\leftarrow \nil$\label{new-leaf-del}
		\State $\x{result} \leftarrow \CAS(\x{Leaf}[k].\x{version}, \x{old}, \x{new})$\label{cas-del}
	\EndIf
	\State $\op{Propagate}(\x{Leaf}[k].\x{parent})$\label{prop-del}
	\State\Return \x{result}
\EndFunction

\medskip

\Function{Boolean}{Refresh}{Node* $x$} \Comment{Try to propagate information to Node $x$ from its children}
	\State $\x{old} \leftarrow x.\x{version}$\label{read-old}
	\State $v_L \leftarrow x.\x{left}.\x{version}$\label{read-left}
	\State $v_R \leftarrow x.\x{right}.\x{version}$\label{read-right}
	\State $\x{new} \leftarrow$ new Version with $\x{left}\leftarrow v_L$, $\x{right}\leftarrow v_R$, $\x{sum}\leftarrow v_L.\x{sum}+v_R.\x{sum}$\label{new-node}
	\State \Return $\CAS(x.\x{version}, \x{old}, \x{new})$\label{cas-refresh}
\EndFunction

\medskip

\Function{}{Propagate}{Node* $x$} \Comment{Propagate updates from $x$'s children up to root}\label{prop-start}
	\While{$x$ is not \nil}\label{prop-loop-start}
		\If{not $\op{Refresh}(x)$}\label{refresh-1}
			\State $\op{Refresh}(x)$ \Comment{Do a second \op{Refresh} if first one fails}\label{refresh-2}
		\EndIf
		\State $x\leftarrow x.\x{paren}t$ \label{go-up}
	\EndWhile\label{prop-loop-end}
\EndFunction

\medskip

\Function{Boolean}{Find}{Key $k$} \Comment{Check if key $k$ is in $S$}
	\State $v\leftarrow \x{Root}.\x{version}$ \Comment{Start at the root}\label{read-root-find}
	\For{$i\leftarrow 1..\log_2 N$} \Comment{Traverse path to leaf of Version tree}
		\If{$i$th bit of binary representation of $k$ is 0} $v\leftarrow v.\x{left}$
		\Else\ $v\leftarrow v.\x{right}$
		\EndIf
	\EndFor
	\State \Return{($v.\x{sum}=1$)}
\EndFunction

\medskip

\Function{int}{Select}{$j$} \Comment{Return the $j$th smallest element in $S$}
	\State $v\leftarrow \x{Root}.\x{version}$ \Comment{Start at the root}\label{read-root-select}
	\State $i\leftarrow 1$ \Comment{Keep track of breadth-first index of $v$ in tree}
	\If{$v.\x{sum} < j$} \Return \nil \Comment{No such element in $S$}
	\Else
		\While{$v.\x{left} \neq \nil$}
			\If{$v.\x{left}.\x{sum} \geq j$} \Comment{Required element is in left subtree}
				\State $v\leftarrow v.\x{left}$
				\State $i\leftarrow 2i$
			\Else \Comment{Required element is in right subtree}
				\State $v\leftarrow v.\x{right}$
				\State $i\leftarrow 2i+1$
				\State $j\leftarrow j-v.\x{left}.\x{sum}$ \Comment{Adjust rank of element being searched for}
			\EndIf
		\EndWhile
		\State \Return $i-N$ \Comment{Convert breadth-first index to value}
	\EndIf	
\EndFunction
\end{algorithmic}
\caption{Implementation of wait-free augmented trie.}\label{pseudocode}
\end{figure}

\begin{figure}
%\begin{subfigure}[t]{0.45\textwidth}
%\input{insert3-1.pdf_t}
%\caption{Initial state when the set is empty.}
%\end{subfigure}
%\hfill
\begin{subfigure}[t]{0.3\textwidth}
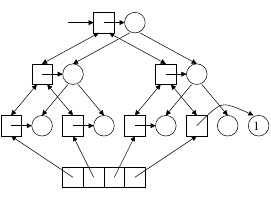
\caption{After the \op{CAS} on line \ref{cas-ins} updates the \x{version} of the rightmost leaf Node.\label{insert3-a}}
\end{subfigure}
\hfill
\begin{subfigure}[t]{0.3\textwidth}
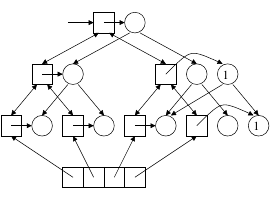
\caption{After the \op{CAS} on line \ref{cas-refresh} updates the right child of the root Node in the first iteration of the loop in \op{Propagate}.\label{insert3-b}}
\end{subfigure}
\hfill
\begin{subfigure}[t]{0.3\textwidth}
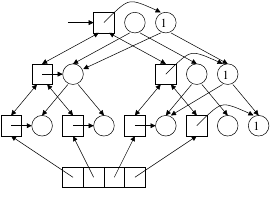
\caption{After the \op{CAS} on line \ref{cas-refresh} updates the root Node in the second iteration of the loop in \op{Propagate}.\label{insert3-c}}
\end{subfigure}
\caption{Key steps of an \op{Insert}($3$) into the initially empty set shown in \Cref{init-fig}.\label{insert3-fig}}
\end{figure}

A \boldop{Refresh}($x$) reads the \x{version} field
of $x$ and its two children, creates a new Version for $x$ based on information in the children's 
Versions, and then attempts to \CAS\ the new Version into $x.\x{version}$.
To handle different augmentations, one must only change the way \op{Refresh} computes the new fields.
\boldop{Propagate}($x$) %(lines \ref{prop-start}--\ref{prop-loop-end}) 
performs a double \op{Refresh} at each node along the path from $x$ to the root. %, as described above.

An \boldop{Insert}($k$) first checks if the key $k$ is already present in the set 
at line \ref{test-ins}.  
If not, it uses a \op{CAS} at line \ref{cas-ins} to change the leaf's
Version object to a new Version object with \x{sum} field 1.
If the \op{CAS} succeeds, the \op{Insert} will return \true\ to indicate a successful insertion.
If the key $k$ is already present when the read at line \ref{read-old-ins} is performed
or if the \op{CAS} fails (meaning that some concurrent operation has already inserted $k$),
the \op{Insert} will return false.
In all cases, the \op{Insert} calls \op{Propagate} before returning to ensure that the information
in the leaf's Version is propagated all the way to the root.

The \boldop{Delete}($k$) operation is very similar to an insertion, except that the operation
attempts to switch the \x{sum} field of $\x{Leaf}[k]$ from 1 to 0.

\boldop{Find} and \boldop{Select} are given as examples of  query operations  in \Cref{pseudocode}.
Each first takes a snapshot of the Version tree by reading $\x{Root}.\x{version}$ on line \ref{read-root-find} or \ref{read-root-select} and then
executes the query's standard sequential code on that tree.
Other queries can be done similarly.  
In particular, to ensure linearizability, queries should access the tree only via $\x{Root}.\x{version}$, not
through the \x{Leaf} array.

\subsection{Correctness}
\label{static-correct}

A detailed proof of correctness appears in \Cref{static-proof}; we sketch it here, with references to claims that are formalized in the appendix.
We first look at the structure of Version trees.
Let $U_x$ be the sequence of keys from the universe $U$ that are represented in the subtree 
rooted at Node $x$ of the tree, in the order they appear from left to right.
In particular, $U_{\x{Root}}=\langle 0,1,\ldots,N-1\rangle$.
We prove (in \Cref{shape}), by induction on the height of the Node $x$, that the 
Version tree rooted at $x.\x{version}$ is a perfect binary tree with $|U_x|$ leaves.
Recall that the fields of Version objects are immutable, so we need only consider
lines \ref{new-leaf-ins}, \ref{new-leaf-del} and \ref{new-node}, which create new Version objects.
\Cref{shape} can be easily proved 
because of the way the Version tree for $x$ is constructed at line \ref{new-node}
by combining the Version trees for $x$'s children.
Line \ref{new-node} also ensures that, for every internal Version $v$, 
$v.\x{sum} = v.\x{left}.\x{sum} + v.\x{right}.\x{sum}$ (\Cref{sums-correct}). 
Since leaf Versions contain 0 or 1, according to lines \ref{new-leaf-ins} 
and \ref{new-leaf-del}, so $v.\x{sum}$ stores the sum of the bits 
stored in leaves in the subtree rooted at $v$.

The key goal of the correctness proof is to define linearization points for the update operations (insertions and deletions)
so that, at all times, the Version tree rooted at $\x{Root}.\x{version}$
 accurately reflects all update operations linearized so far.
Then, we linearize each query operation at the time it reads $\x{Root}.\x{version}$ to take
a snapshot of the Version tree.  This will ensure that the result returned by the query is consistent
with the state of the represented set $S$ at the query's linearization point.

We consider an arbitrary execution in which processes perform operations on the trie.
An execution is formalized as an alternating sequence of configurations and steps
$C_0, s_1, C_1, s_2, \ldots$, where each configuration $C_i$ describes the state of 
the shared memory and the local state of each process, and each $s_i$ is a step by some 
process that takes the system
from configuration $C_{i-1}$ to $C_i$.  A step is either a shared-memory access or a local step
that affects only the process's local state.

Our goal is to define a linearization point (at a step of the execution) of each update operation so that for each configuration $C$,
the Version tree rooted at $\x{Root}.\x{version}$ is the trie that would result by sequentially performing
all the operations that are linearized before $C$ in their linearization order.
Thus, the linearization point of an update operation should be the moment when the effect of the update
has been propagated to the root Node, so that it becomes visible to queries.  
To define these linearization points precisely,
we define the \emph{arrival point} of an update operation on a key $k$ at each Node along
the path from the leaf Node representing $k$ up to the root Node.
Intuitively, the arrival point of the update at Node $x$
is the moment when the effect of the update is reflected in the 
Version tree rooted at $x.\x{version}$.
Then, the linearization point is simply the arrival point of the update at \x{Root}.
We must ensure these linearization points are well defined by showing 
that the double-refresh technique propagates each update all the way up to \x{Root} before
the update terminates.

\Cref{arrival}, below, formally defines the arrival point of each $\op{Insert}(k)$ or $\op{Delete}(k)$ operation at Node $x$, 
where $k\in U_x$ using induction from the bottom of the tree to the top.
If an \op{Insert}($k$) sees that $k$ is already in a leaf Node at line \ref{read-old-ins} , or if a \op{Delete}($k$) sees that $k$ is not present in a leaf Node at line \ref{read-old-del}, the arrival point of the operation is at that line.
Otherwise the update performs a \CAS\ on the leaf at line \ref{cas-ins} or \ref{cas-del}.
If the \CAS\ succeeds, the \CAS\ is the update's arrival point at that leaf.
Otherwise, we put the arrival point of the update at the leaf 
at a time when $k$'s presence or absence would cause the update to fail.
An update's arrival point at an internal Node is the first successful \CAS\ by a \op{Refresh} that
previously read the child after the update's arrival point at that child.

\begin{definition}
\label{arrival}
We first define the arrival point of an $\op{Insert}(k)$ or $\op{Delete}(k)$ operation $op$ at $\x{Leaf}[k]$.
\begin{enumerate}
\item\label[part]{arr-leaf-suc}
If $op$ performs a successful \CAS\ at line \ref{cas-ins} or \ref{cas-del}, then the arrival point of $op$ is that \CAS.
\item\label[part]{arr-leaf-fail}
If $op$ performs an unsuccessful \CAS\ at line \ref{cas-ins} or \ref{cas-del}, then the arrival point of $op$ is the first successful \CAS\ on $\x{Leaf}[k].\x{version}$ after $op$ read the old value of $\x{Leaf}[k].\x{version}$ at line \ref{read-old-ins} or \ref{read-old-del}.  (Such a \CAS\ must exist; otherwise $op$'s \CAS\ would have succeeded.)
\item\label[part]{arr-leaf-read}
If $op$ is an \op{Insert} that reads a Version with $\x{sum}=1$  from $\x{Leaf}[k].\x{version}$ on line \ref{read-old-ins}
or $op$ is a \op{Delete} that reads a Version with $\x{sum}=0$ from $\x{Leaf}[k].\x{version}$ on line \ref{read-old-del},
then the arrival point of $op$ is $op$'s read at line \ref{read-old-ins} or \ref{read-old-del}.
\end{enumerate}
If multiple operations' arrival points at a leaf Node are at the same successful \CAS, we order them:  first the operation that did
the successful \CAS, then all the other operations (ordered arbitrarily).

Next, we define the arrival point of an $\op{Insert}(k)$ or $\op{Delete}(k)$  $op$ at an internal Node $x$ with~$k\in U_x$.
\begin{enumerate}
\setcounter{enumi}{3}
\item\label[part]{arr-left}
If $k\in U_{x.\x{left}}$, the arrival point of $op$ is the first successful \CAS\ on
$x.\x{version}$ at line \ref{cas-refresh} of a \op{Refresh} that read $x.\x{left}.\x{version}$ at line \ref{read-left} after the arrival point of $op$ at $x.\x{left}$.
\item\label[part]{arr-right}
If $k\in U_{x.\x{right}}$, the arrival point of $op$ is the first successful \CAS\ on
$x.\x{version}$ at line \ref{cas-refresh} of a \op{Refresh} that read $x.\x{right}.\x{version}$ at line \ref{read-right} after the arrival point of $op$ at $x.\x{right}$.
\end{enumerate}
If multiple operations' arrival points at an internal Node are at the same successful \CAS, we order them as follows:  first the operations on keys in $U_{x.\x{left}}$ in the order they arrived at $x.\x{left}$ and then the operations on keys in $U_{x.\x{right}}$ in the order they arrived at $x.\x{right}$.
\end{definition}
%\here{The definition can become shorter (e.g. by combining cases 4 and 5 to save space.}

For example, consider the \op{Insert}(3)  depicted in \Cref{insert3-fig}.
Its arrival point at the leaf Node representing key 3 is the CAS that updates
that leaf's \x{version} field, shown in \Cref{insert3-a}.
Its arrival point at the parent of this leaf is the CAS that updates the data
structure as shown in \Cref{insert3-b}.
Its arrival point at the root is the \CAS\ that updates the \x{Root}.\x{version} as shown in \Cref{insert3-c}.

\begin{figure}
\begin{minipage}[b]{0.4\linewidth}
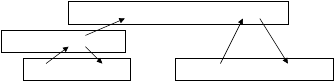
\caption{Call to \op{Refresh} in proof that a double refresh successfully propagates updates to a Node from its children.  The horizontal axis represents time, and boxes indicate the interval  between a routine's invocation and its response.  
Numbers refer to line numbers in the pseudocode.
An arrow $s_1 \rightarrow s_2$ indicates step 
$s_1$ must precede step $s_2$.\label{before-end-fig}}
\end{minipage}
\hfill
\begin{minipage}[b]{0.55\linewidth}
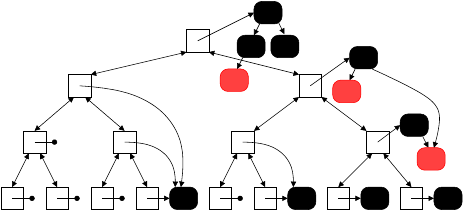
\caption{Augmenting the trie with red-black trees (RBTs) to speed up queries.  $N=8$ and $S=\{3,5,6,7\}$.  Squares are trie Nodes.  Ovals are RBT nodes.  Each RBT node has child pointers, and stores a key and a $size$ field that represents the number of keys in the subtree.  Black dots represent RBT nodes with $sum$~0.\label{trie-rbt}}
\end{minipage}%
\end{figure}

It follows easily from \Cref{arrival} that arrival points of an update operation $op$
are after $op$ begins.  (See \Cref{after-start}.)
If $op$ terminates, we also show (in \Cref{before-end}) that it has an arrival
point at the root Node before it terminates.  
Recall that after $op$'s arrival point at a leaf, $op$ calls \op{Propagate}, which does a double \op{Refresh} at each Node
along the path from that leaf to the root.
%As in proofs for previous algorithms that use the double refresh technique,
We show by induction that the double refresh at each node $x$ along the path
ensures $op$ has an arrival point at $x$.
The induction step follows immediately from \Cref{arr-left,arr-right} of \Cref{arrival}
if one of $op$'s calls to \op{Refresh}($x$) performs a successful \CAS.
So, suppose both of $x$'s calls $R_1$ and $R_2$ to \op{Refresh}($x$) fail their \CAS.
Then for each $R_i$, there must be a successful \CAS\ $c_i$ on $x.\x{version}$ between 
$R_i$'s read of $x.\x{version}$ on line \ref{read-old} and its \CAS\ 
on line \ref{cas-refresh}, as depicted in \Cref{before-end-fig}.
Although $c_1$ may store outdated information, the \op{Refresh} that performs
$c_2$ must have read information from $x$'s children after $c_1$, which is enough
to ensure that $op$ has an arrival point at $x$, by \Cref{arr-left,arr-right} of \Cref{arrival}.%

Our next goal is to prove a key invariant that, 
for each configuration $C$ and Node $x$, 
the Version tree rooted at $x.\x{version}$ accurately reflects all of the updates whose
arrival points at $x$ are prior to $C$.
In other words, it is a trie structure (similar to the one shown in \Cref{bitvector})
that would result from performing all of those updates in the order of their
arrival points at $x$.
As a corollary, when we take $x$ to be the root Node,
we see that the Version tree rooted at $\x{Root}.\x{version}$ has a 1 in the leaf
for key $k$ if and only if $k$ is in the set obtained by sequentially performing 
the linearized operations in order.  Correctness of all query operations
follows from this fact and the fact that \x{sum} fields in Version trees are accurate (\Cref{sums-correct}, described above).

We sketch the proof of the key invariant.
We make the argument separately for each key $k \in U_x$.
We define $Ops(C,x,k)$ to be the sequence of update operations on key $k$ whose
arrival points at $x$ precede configuration $C$, in the order of their arrival points.
We must show that, in each configuration~$C$,
the leaf corresponding to key $k$ in the subtree rooted at $x.\x{version}$
contains a 1 if and only if $Ops(C,x,k)$ ends with an \op{Insert}($k$).
(See \Cref{inv:correct}.)

If $x$ is the leaf for key $k$, we consider each step that can add arrival points at $x$.
First, consider a \CAS\ that flips the bit stored in $x.\x{version}$.
If the \CAS\ sets the bit to 1, it follows from \Cref{arr-leaf-suc} and \Cref{arr-leaf-fail} of
\Cref{arrival}
that it is the arrival point of one or more
\op{Insert}($k$) operations, which preserves the invariant.  Similarly, a \CAS\ that sets the bit to 0 is the
arrival point of one or more \op{Delete}($k$) operations, which preserves the invariant.
If the step is
%The other step where arrival points at $x$ may be defined is 
an \op{Insert}($k$)'s read of $x.\x{version}$ when it has value 1 or a \op{Delete}($k$)'s read of $x.\x{version}$ when it has value 0, it also preserves the invariant.

If $x$ is an internal Node, the fact that the invariant holds at $x$ can be proved inductively.
The claim at $x$ follows from the assumption that it holds at the children of $x$, since the invariant is phrased in terms
of a single key and the sets of keys represented in the two subtrees of $x$ are disjoint.

Finally, we prove that operations that arrive at a leaf
are propagated up the tree in an orderly way,
so that they arrive at the root in the same order.
This is useful for showing that the update operations return results consistent with their
linearization order (\Cref{updates-correct}).

\subsection{Complexity and Optimizations}

\op{Insert} and \op{Delete} take $O(\log N)$ steps. 
Searches and the order-statistic queries listed at the beginning of Section \ref{static} 
take $O(\log N)$ steps and are read-only.
%(This contrasts with the non-concurrent version
%of this data structure, where the answer to a query can simply be read from the appropriate
%leaf in $O(1)$ time.)
\op{Size} queries can be answered in $O(1)$ time by simply returning $\x{Root}.\x{version}.\x{sum}$.
We could also augment the data structure so that each node stores the minimum element in its subtree
to answer \op{Minimum} queries in $O(1)$ time.
A range query that returns $R$ elements can be done in $O(R(\log\frac{N}{R}+1))$ time,
since it visits at most $R$ locations in the top $\log R$ levels of the Version tree
and in the rest of the tree it visits $O(\log N - \log R)$ locations per returned element, for a total of $O(R(\log N-\log R+1))$ locations.
All operations are wait-free.

%\here{I thought that this data structure might circumvent the lower bound of Attiya and Fouren \cite{AF17}
%for bag data structures
%using the assumption of a bounded universe, but that seems not to be the case, because I don't see
%how to implement a bag's remove operation without running into the CAS retry problem.}

%\here{Should be possible to prove that the space usage is $O(pN)$ where $p$ is the number of processes.
%I.e., at any time the amount of memory that cannot be freed because it is reachable at that time is $O(pN)$.
%Or if there are currently u updates and q queries pending, is it possible to prove that the reachable space is $O(N+uN+q\log N)$?
%}

%\subsection{Optimizations}

We assume a safe garbage collector, such as the one provided by Java, which deallocates
objects only when they are no longer reachable.
%It deallocates a Version only when that Version cannot be reached by following pointers
%from any Node or from a Version that some process holds a pointer to.
%Old Versions are automatically unlinked from our data structure when they are replaced by new Versions.
%once a Version is no longer reachable from the root, it will remain reachable only as long
%as some query operation is working on an old Version tree that contains it.
%If the new Version created on line \ref{new-leaf-ins}, \ref{new-leaf-del} or \ref{new-node} is not installed by the
%subsequent \CAS\ on line \ref{cas-ins}, \ref{cas-del} or \ref{cas-refresh}, it can be deallocated immediately.  Or,
%to reduce the number of allocations and deallocations,
%the unused Version can be kept and reused the next time the process needs to create a new Version.
%(Although this violates the immutability of fields of a Version, it does not cause any problems,
%because the fields of a Node remain unchanged once a pointer to it is written into shared memory, which is the important
%property for correctness.)
We now consider a very pessimistic worst-case bound on the amount of space 
used by objects that are still reachable. % (and therefore cannot be garbage collected).
For each Node $x$, up to $O(\log N)$ different Versions belonging to $x$ could be stored in
the Version trees of each of $x$'s ancestors.
Thus, the space used by all objects reachable by following pointers from $Root$ 
is $O(N\log N)$.
In addition to this, any old ongoing queries could have an old snapshot of a Version
tree.

The Node tree is static and complete, so it can be represented using
an array $\x{Tree}[1..2N-1]$ of pointers to Versions, where $\x{Tree}[1]$ is  the root, and the children of the
internal Node $\x{Tree}[i]$ are  $\x{Tree}[2i]$ and $\x{Tree}[2i+1]$ \cite[p.\ 144]{TAOCP-3}.
This %representation 
saves the space needed for the \x{Leaf} array and parent and child pointers, since we can navigate the tree by index arithmetic rather than following pointers.
%The \x{Leaf} array is also unnecessary using this representation since $\x{Leaf}[i]$ is simply $\x{Tree}[i+N]$.
%Version objects would still store pointers to their children.
%The \x{Leaf} array is also unnecessary using this representation. % since $\x{Leaf}[i]$ is simply $\x{Tree}[i+N]$.
%Version objects would still store pointers to their children.

%\here{To think about:  might there be a way of detecting when your operation has been helped to arrive at the root
%so that you don't have to continue working on the operation yourself?  Might save some CAS steps.  But this seems
%difficult.}

\subsection{Variants and Other Applications}
\label{variants}

We described how to augment the trie with a $sum$ field to facilitate efficient order-statistic queries.
However, the method can be used for any augmentation where the values of a node's additional
fields can be computed from information in the node and its children, 
by modifying line \ref{new-node} to compute the new fields.
%as in the
%general technique described by Cormen et al.~\cite{CLRS4-17}.
\Cref{introduction} mentions some of the many applications where this can be applied.
%bit vectors with rank and select queries are used a lot in the design of compact/succinct data structures, so this might be a step towards developing concurrent implementations of some of those data structures.

Generalizing our implementation to $d$-ary trees is straightforward for any $d\geq 2$.  The number of CAS instructions per update would be reduced
to $2\log_d N$, but the number of reads (and local work) per update 
would increase to $\Theta(d\log_d N)$.  Order statistic queries could still
be made to run in $\Theta(\log_2 N)$ time if each node stores prefix sums and uses binary search.

Instead of storing a set of keys $S\subseteq U$, a straightforward variant of our data structure can store a set of key-value pairs where each record has a unique 
key drawn from $U$.
Instead of storing just one bit,
a leaf's Version object would also store the associated value.
A \op{Replace}$(k,v)$ operation 
that replaces the value associated with key $k$ with a new value $v$ would update the appropriate leaf's $version$ field and call \op{Propagate}.
If several \op{Replace}($k,*$) operations try to update a leaf concurrently, 
one's \op{CAS} will succeed and the others will fail, and we can assign them all arrival points at the leaf at the time of the successful CAS, with the
failed operations preceding the successful one.

Our approach can also provide lock-free \emph{multisets}
of keys drawn from $U$.
Instead of storing a bit, the leaf
for key $k$ stores a Version whose \x{sum} field is the number of copies of $k$ in the multiset.
With \CAS\ instructions, operations can be made lock-free
if each \op{Insert}($k$) or \op{Delete}($k$) repeatedly tries to
install a new Version $k$'s leaf with its \x{sum} field incremented or decremented and
then calls \op{Propagate}.
If the leaf's \x{sum} field can be updated with a fetch\&add, the updates can be made wait-free.
%It should be possible to show that the amortized time complexity of this operation is $O(c+\log N)$, where $c$ is point contention.

Without any modification, our trie supports multipoint
queries, like range searches that return all keys in a given range,
since reading $Root.version$ yields a snapshot of the trie.
In fact, our technique has more efficient queries than some recent
papers discussed in Section \ref{related} that provide multipoint queries:
in our approach, queries take the same number of steps as in a sequential implementation.

\subsection{Improving Query Time to $O(\log n)$}
\label{faster-queries}

The time to perform order-statistic queries on the set $S$ can be improved from $O(\log N)$ 
to $O(\log |S|)$.
To do this, we simply use a different augmentation.
The \x{version} field of each Node $x$
stores a pointer to the root of a red-black tree (RBT) that represents
all the elements in the subtree of Nodes rooted at $x$.
See \Cref{trie-rbt} for an example.
A \op{Refresh}$(x)$ updates $x.\x{version}$ by reading the RBTs stored
in $x.\x{left}.\x{version}$ and $x.\x{right}.\x{version}$, joining them into one RBT
(without destroying the smaller RBTs) and then using a CAS to store
the root of the joined RBT in $x.\x{version}$.
The algorithm to \op{Join} two RBTs in logarithmic time, provided that all elements in one are smaller than
all elements in the other, is in Tarjan's textbook \cite{Tar83}.
To avoid destroying the smaller RBTs when performing a \op{Join}, one
can use the path-copying technique of Driscoll et al.~\cite{DSST89}.
Pseudocode is  in \Cref{faster-queries-app}.

Each RBT node also has a $size$ field storing the number of elements
in the subtree rooted at that node.
A query reads \x{Root}.\x{version} to get
a snapshot of a RBT containing all elements in the dynamic set.
Order-statistic queries are answered in $O(\log n)$ steps using the \x{size} fields of the RBT.

There is a tradeoff:  the time for updates increases to 
$O(\log N\log \hat{n})$, since a \op{Join} of two RBTs must be performed at each 
of $\log N$ Nodes of the Node tree during \op{Propagate}.
Here, $\hat{n}$ denotes a bound on the maximum size the set could have
under any possible linearization of the update operations.
The elements in a RBT constructed by a \op{Refresh} on a non-root Node may
never all be in the set simultaneously, so we must argue that the size
of each such RBT is $O(\hat{n})$.
Consider a \op{Join}($T_1,T_2$)  during a call $R$ to \op{Refresh}($x$).
Without loss of generality, assume $|T_1|\geq |T_2|$.
Let $\alpha'$ be the prefix of the execution up the time $R$
reads $T_1$ from $x.\x{left}.\x{version}$.
Suppose we modify $\alpha'$ by delaying $R$'s read of $x.\x{version}$ 
until just before $R$ reads $x.\x{left}.\x{version}$, and then appending
to the execution all the steps needed to complete the \op{Propagate} that called $R$.
This will ensure that all remaining \CAS\ steps of the \op{Propagate} succeed
and $T_1$ will be a subtree of the tree stored in \x{Root}.\x{version}.
Thus, there must be some way to linearize  $\alpha'$ so that 
all elements in $T_1$ are simultaneously in the represented set (since the modified execution is linearizable), so $|T_1|\leq \hat{n}$.
Thus, the size of the RBT that $R$ builds is $|T_1|+|T_2| \leq 2|T_1| \leq 2\hat{n}$.

% !TEX root =  augmented-tree.tex

\section{Augmented Binary Search Tree}
\label{bst-sec}

\begin{figure}
\begin{center}
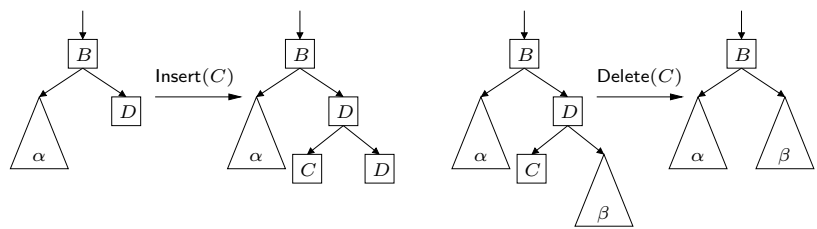
\end{center}
\caption{How updates modify a leaf-oriented BST.  Here, $\alpha$ and $\beta$ represent arbitrary subtrees.\label{tree-changes}}
\end{figure}

In this section, we illustrate our technique by augmenting a binary search tree (BST) that represents
a set $S$ of elements drawn from an {\em arbitrary} (ordered) universe $U$. 
We describe the augmentation for order-statistic queries, but as explained above,
the same approach can be used for many other applications. 
In constrast to the augmented trie of \Cref{static}, 
the time and space complexity of our augmented BST depend on $|S|$ rather than $|U|$.
%Although the technique for augmenting the BST is quite similar to the trie,
%the proof of correctness is complicated by the fact that the shape of the tree 
%changes as elements are inserted and deleted.

\subsection{Basic Lock-free BST}
\label{basic-bst-sec}

We base our augmented BST on the lock-free BST of Ellen et al.~\cite{EFHR14}, so
we first give a brief overview of how their BST works.
The BST is leaf-oriented:  keys of $S$ are stored in the leaves; keys in internal nodes
serve only to direct searches to the leaves.
%  (This design decision simplifies deletions.)
The BST property requires that all keys in the left subtree of a node $x$ are smaller than $x$'s key
and all keys in the right subtree of $x$ are greater than or equal to $x$'s key.
The tree nodes maintain child pointers, but not parent pointers.
To simplify updates, the BST is initialized with three sentinel nodes: an internal node
and two leaves containing dummy keys $\infty_1$ and $\infty_2$, 
which are considered greater than any actual key in $U$ and are never deleted.
A shared $Root$ pointer points to the root node of the tree, which never changes.

An insert or delete operation starts at the root and searches for the leaf at which to apply its update.
Updates are accomplished by simple modifications to the tree structure as shown in \Cref{tree-changes}.
To coordinate concurrent updates to the same part of the tree, updates must flag a node before modifying one of
its child pointers and remove the flag when the modification is done.
Before removing an internal node from the tree, the operation must \emph{permanently} flag it.
Since only one operation can flag a node at a time, flagging a node is analogous to locking it.
To ensure lock-free progress, an update that needs to flag a node that is already
flagged for another update first \emph{helps} the other update to complete and then tries again to perform its operation.
When retrying, the update does not begin all over from the top of the tree;
the update keeps track of the sequence of nodes it visited on a thread-local stack so 
that it can backtrack a few steps up the tree by popping the stack
until reaching a node that is not
permanently flagged for deletion, and then searches onward from there for the location
to retry its update.
%Further details of this coordination scheme are not essential for understanding how to augment the tree.
Each update is linearized at the moment
one of the changes shown in \Cref{tree-changes} is made to the tree, 
either by the operation itself or by a  helper.

The tree satisfies the BST property at all times.  We define the \emph{search path} for a key $k$ at some configuration $C$ to be the
path that a sequential search for $k$ would take if it were executed without interruption in $C$.
Searches in the lock-free BST ignore flags and simply follow child pointers until reaching a leaf.
A search for  $k$ may pass through nodes that get removed by concurrent updates,
but it was proved in \cite{EFHR14} that each Node the search visits
\emph{was} on the search path for $k$ (and by the way we linearize updates, it was thus also in the set represented by the BST)
at some time during the search. 
A search that reaches a leaf $\ell$ is linearized when that leaf was on the search path for $k$.

\subsection{Lock-free Augmentation}

\begin{figure}
\begin{subfigure}[t]{0.23\textwidth}
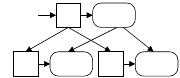
\caption{Initialization of augmented BST with\linebreak sentinel Nodes.\label{bst-init}}
\end{subfigure}
\hfill
\begin{subfigure}[t]{0.73\textwidth}
\hspace*{4mm}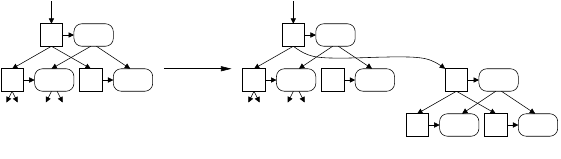
\caption{Addition of new Nodes and Versions for an \op{Insert}.    In this example, the subtree rooted at $A$ has four leaves.  The data structure is shown after the three new Nodes have been added to the Node tree, but before the change has been propagated to Node $B$'s Version.\label{succ-ins-fig}}
\end{subfigure}

\begin{subfigure}[t]{\textwidth}
\hspace*{27mm}
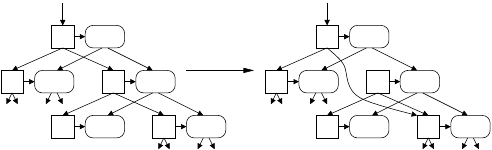
\caption{Change to the Node tree for a \op{Delete}.
The subtrees rooted at $A$ and $E$ have four and three leaves, respectively.  The data structure
is shown after the Nodes $C$ and $D$ have been removed from the Node tree, but before the
change has been propagated to Node $B$'s Version.\label{succ-del-fig}}
\end{subfigure}
\caption{Augmented BST data structure.
Nodes are shown as squares and Version objects as ovals with $key$ and  \x{sum} fields shown.}
\end{figure}

We now describe how to augment the lock-free BST of \cite{EFHR14} 
with additional fields for each node, provided the fields can
be computed from information in the node and its children.
We again use the \x{sum} field, which supports efficient order-statistic queries, as an illustrative example.
As in \Cref{static-alg}, we add to each tree Node $x$ a new \x{version} field 
that stores a pointer to a tree of Version objects.  This Version tree's leaves
form a snapshot of the portion of $S$ stored in the subtree rooted at $x$.
In particular, the leaves of the Version tree stored in $\x{Root}.\x{version}$ 
form a snapshot of the entire set~$S$.
Each Version  $v$ stores a \x{sum} field and pointers
to the Versions of $x$'s children that were used to compute $v$'s \x{sum}.
Each Version associated with Node $x$ also stores a copy of $x$'s key
to direct searches through the Version trees.
Version trees will always satisfy the BST property, and the \x{sum} field of each Version $v$
stores the number of keys in leaf descendants of $v$.
See \Cref{bst-types} on page \pageref{bst-types} for a formal description of the Node and Version object types.
See \Cref{bst-init} for the initial state of the BST, including the sentinel Nodes.
Pseudocode for the implementation is  in \Cref{BST-pseudo-section}.

An \boldop{Insert} or \boldop{Delete} first runs the algorithm from \cite{EFHR14} to modify the Node tree as shown in \Cref{tree-changes}.
\Cref{succ-ins-fig,succ-del-fig} show the effects of the modification when Versions are also present.
Then, the update calls \op{Propagate} to modify the
$sum$ fields of the Versions of all Nodes along the path from
the location where the key was inserted or deleted to the root.
As in \Cref{static-alg}, 
an update operation's changes to the $sum$ field of all thes Nodes become
visible at the same time, and we linearize the update at that time.
%We use a scheme similar to that of \Cref{static-alg} to make all of these updates to $sum$ fields become visible at the same time, and that moment is the linearization point of the update.
If an \op{Insert}($k$) reaches a leaf Node that already contains $k$, before returning \false,
it also calls \op{Propagate} to ensure that the operation that inserted the other copy 
of key $k$ has been propagated to the $Root$ (and therefore linearized).
Similarly, a \op{Delete}($k$) that reaches a leaf Node and finds that $k$ is absent from $S$
also calls \op{Propagate} before returning \false.

The \boldop{Propagate} routine is similar to the one in \Cref{static-alg}.
As mentioned in \Cref{basic-bst-sec}, each update uses a thread-local stack to keep track of the Nodes
that it visits on the way from $Root$ to the location where the update must be performed,
so \op{Propagate} can simply pop these Nodes off the stack 
and perform a double \op{Refresh} on each of them.
Some of the Nodes along the path may have been removed from the Node tree by
other \op{Delete} operations that are concurrent with the update, but there is no harm in applying
a double \op{Refresh} to those deleted Nodes.

As in \Cref{static-alg}, each \boldop{Refresh} on a Node $x$ reads the Versions
of $x$'s children and combines the information in them to create a new Version for $x$,
and then attempts to \CAS\ a pointer to that new Version into $x.version$.
There is one difference in the \op{Refresh} routine:  because $x$'s child pointers
may be changed by concurrent updates,
\op{Refresh} reads $x$'s child pointer, reads that child's $version$ field,
and then reads $x$'s child pointer again.  If the child pointer has changed,
\op{Refresh} does the reads again, until it gets a consistent
view of the child pointer and the $version$ field of that child.

A \textbf{query operation} first reads $Root.version$ to get the root of a Version tree.
This Version tree is an immutable BST (with $sum$ fields) whose leaves form
a snapshot of the keys in $S$ at the time $Root.version$ is read.  The query is linearized at this read.
Then, the standard, sequential
algorithm for an order-statistic query can be run on that Version tree.
To ensure linearizability, searches are performed like other queries.
This has the additional benefit of making searches wait-free,  unlike the original BST of \cite{EFHR14}, where searches can starve.
Complex queries, like range queries, can access any subset of Nodes in the snapshot.
Our technique provides snapshots in a  simpler way than
\cite{FPR19} (later generalized by \cite{WBB+21} to any CAS-based data structure),
which keeps a list of previous timestamped versions
of each child pointer.
Our approach makes queries more efficient since they 
do not have to search back through version
lists for an old version with a particular timestamp.  It also avoids many
of the problems of garbage collection, since old Versions are automatically
disconnected from our data structure when a new Version replaces it.
Unlike \cite{FPR19}, our approach does not provide a snapshot of the Node tree:
the shape of the Version tree may not match the shape of the Node tree at any time.
Instead, our approach provides a snapshot of the \emph{set of elements
represented by the tree}.

%\here{If one process helps another complete its update, it does not have to call \op{Propagate}, right?}

\subsection{Correctness}
\label{bst-correct-short}

As in \Cref{static-correct}, we define arrival points of update operations at a Node to keep track of when
the updates have been propagated to that Node.
%, and are reflected in the Version tree stored there.
We again linearize updates at their arrival point at
the root Node, and queries when they obtain a snapshot of the Version tree by 
reading \x{Root}.\x{version}.
As in \cite{EFHR14}, sentinel
Nodes as shown in \Cref{bst-init} 
ensure that the root Node never changes. 

The proof again has two main parts:  showing
that every update operation has an arrival point at the root, % before the update terminates,
and showing an invariant that in every configuration $C$, 
the Version tree rooted at a Node $x$ is a legal (augmented)
BST containing the set that would result from sequentially performing all operations
that have arrival points at $x$ at or before $C$, in the order of their arrival points.
The former claim implies that the linearization respects the real-time order of operations.
%Applying the latter invariant is applied to the root allows us to prove that the 
%queries return results consistent with the linearization ordering.
Applying the latter claim to the root shows that  
queries return results consistent with the linearization ordering.

Although this high-level plan for the proof is similar to \Cref{static-correct},
updates' changes to the Node tree introduce some challenges.
Firstly, we must ensure that updates are not ``lost'' if concurrent
updates remove the Nodes to which they have  propagated.
This involves transferring arrival points from the removed Node $x$ to another Node $x'$, 
and this requires proving a number of claims about the arrival points
that can be present at $x$ and $x'$ to ensure that transferring arrival points from $x$ to $x'$
does not change the set of keys that should be stored in the Version tree of $x'$.
Secondly, in the original, unaugmented BST of \cite{EFHR14}, an \op{Insert}($k$) that reaches a leaf that
already contains $k$ returns \false, but that leaf may no longer be in the tree when the \op{Insert} reaches it, so the linearization point of the \op{Insert}
is retroactively chosen to be some time during the \op{Insert} when that leaf
was present in the tree.
We must do something similar in choosing the arrival point of failed updates at a leaf.
%which complicates the proofs of the main claims about the arrival points of operations.

In this section, we define the arrival points (which in turn defines the linearization)
and  sketch some of the key arguments about them.  A detailed proof of
linearizability appears in \Cref{bst-correct-full}.
For a configuration $C$, let $T_C$ be the Node tree in configuration $C$:  this is the tree
of all Nodes reachable from $Root$ by following child pointers.
Since our augmentation does not affect the way the Node tree is handled,
it follows from \cite{EFHR14} that $T_C$ is always a BST.
The \emph{search path for key $k$ in $C$} is the root-to-leaf path in $T_C$
that a BST search for key $k$ would traverse.
The following intuition guides our definition of arrival points:  
the arrival point of an update operation $op$ on key $k$ at a Node $x$ 
should be the first time when both (a) $x$ is on the search path for $k$ and 
(b) the effect of $op$ is reflected in the Version tree rooted at $x.\x{version}$.
We also ensure that, for any configuration $C$, the Nodes at which
an operation has arrival points defined will be a suffix of the search path for $k$ in~$C$.

A successful \op{Insert}($k$), shown in \Cref{succ-ins-fig}, replaces a leaf $\ell$ containing some key $k'$ by a new
internal Node \x{new} with two children, \x{newLeaf} containing $k$, and $\ell'$, which is a new copy
of $\ell$.
The \CAS\ step that makes this change is 
the arrival point of the \op{Insert}($k$) at $new$ and $newLeaf$, since these Nodes' Version trees
are initialized to contain a leaf with key $k$.
There may also be many operations that had arrival points at  $\ell$ before $\ell$ 
is replaced by $\ell'$ in the Node tree.  For example, there may be an \op{Insert}($k''$) followed by a 
\op{Delete}($k''$) if $\ell$ is the end of the search path for $k''$.  If these operations
%have not yet propagated to the root, we must ensure that they are, so that they can be assigned
have not  propagated to the root, we must ensure that this happens, so that they are linearized:  we do not want to lose the arrival points of these
operations when $\ell$ is removed from the Node tree.
So, we transfer all arrival points of update operations at $\ell$ to \x{new} and the appropriate
child of \x{new} (depending on whether the key of the update is less than \x{new}.\x{key} or not).

Similarly, when a \op{Delete}($k$) changes the Node tree as shown in \Cref{succ-del-fig},
each operation with an arrival point at the deleted leaf $\ell$ (and the \op{Delete}($k$) itself)
is assigned an arrival point at
$\ell$'s sibling \x{sib}, and at all of \x{sib}'s descendants on the search path for the operation's key.  
We shall show that operation's key cannot appear
in the Version trees of any of those Nodes, so the Version trees of those Nodes correctly reflect the fact
that the key has been deleted.

As mentioned above, if an \op{Insert}($k$) returns \false\ because it finds 
a leaf $\ell$ containing $k$ in the tree, it was shown in \cite{EFHR14} that $\ell$ \emph{was} on the search path for $k$ in some configuration $C$ during the \op{Insert}.
Since augmentation has no effect on updates' accesses to the Node,
this is still true for the augmented BST.
We choose $C$ as the arrival point of the \op{Insert} at that leaf.
Similarly, if a \op{Delete}($k$) returns \false\ because its search for $k$ reaches a leaf $\ell$
that does not contain $k$, 
we choose a configuration $C$ during the \op{Delete} when $\ell$ is on the search path for $k$ in $T_C$ as the arrival point of the \op{Delete} at $\ell$.

When a \op{Refresh} updates the \x{version} field of a Node $x$, 
we assign arrival points to all update operations that had arrival points
at $x$'s children before the \op{Refresh} read the \x{version} fields of those children,
 as in \Cref{static-correct}.  This indicates that those
operations have now propagated to $x$, and the Version tree in $x.version$
reflects the fact that those updates have been performed.

\Cref{arrival-bst} in \Cref{bst-correct-full} formalizes the arrival points described above.
We then use this definition to prove that each update operation's arrival point
at the root is between the update's invocation and response (refer to \Cref{bst-order-proof}).  In particular, this reasoning has to argue that no operation
gets ``lost'' as it is being propagated to the root if concurrent deletions remove
Nodes to which it has been propagated.
Recall that \op{Propagate} calls a double \op{Refresh} on every Node in the update operation's
local $stack$, which remembers all of the internal Nodes visited to reach the leaf $\ell$
where the update occurs.
We use the fact from \cite{EFHR14} that Nodes can be removed from the path that leads from the root
to $\ell$, but no new Nodes can ever be added to it.  (It is fairly easy to see that 
the changes to the Node tree shown in \Cref{tree-changes} cannot add a new ancestor to $\ell$.)
Thus, \op{Propagate} calls a double \op{Refresh} on every ancestor of $\ell$ to propagate
the update all the way to the root Node.

Then, we prove our main invariant that in every configuration $C$ and in each Node $x\in T_C$,
the leaves of the Version tree stored in $x.\x{version}$ contain exactly those
keys that would be obtained by sequentially performing the operations with arrival points at $x$ at or before $C$, in the order of their arrival points
(see \Cref{accurate} in \Cref{bst-match-proof}.).
%For the BST, 
This argument is complicated by the fact that the Node tree changes
and arrival points are shifted from one Node to another.  We make the argument
by focusing on one key $k$ at a time, and showing that $k$ is in the Version tree 
if and only if the last operation on key $k$ in the sequential execution is an \op{Insert}.
We also show that the boolean responses these operations will return are consistent with this sequential ordering.
Applying this invariant  to the root Node, this implies updates
return responses consistent with the linearization ordering.

Unlike the trie in \Cref{static}, the subtree rooted at Node $x$ may have a different
shape than the Version tree rooted at $x.\x{version}$, since updates may have
changed the Node tree since $x.\x{version}$ was stored.
However, we prove (in \Cref{left-right}) that for any configuration $C$ and any Node $x\in T_C$,
the keys of update operations with arrival points at Nodes in the left (or right) subtree of $x$
are less than $x.\x{key}$ (or greater than or equal to $x.\x{key}$, respectively).
Together with the main invariant mentioned above, this allows us to prove that 
all Version trees are legal BSTs.  The correctness of the augmentation fields is trivial,
since these fields are correct when an internal Version is created, and its fields are immutable.
It follows that  queries  return results consistent with the linearization.

\subsection{Complexity}

The amortized time per operation on the unaugmented BST
is $O(h+c)$, where $h$ is the height of the Node tree and $c$ is point contention \cite{EFHR14}.
Since we have not made any change to the way the Node tree is handled,
we must just count the additional steps required for the augmentation.
We argue that the amortized time to perform a \op{Propagate} is also $O(h+c)$.
The number of iterations of the loop in \op{Propagate}
is bounded by the number of elements pushed on to the stack by the update,
which in turn is bounded by the running time of the update in the original algorithm of \cite{EFHR14}.
Recall that a \op{Refresh} may have to reread child pointers repeatedly until it gets
a consistent view of the child pointer and the child's $version$ field.
Rereading is necessary only if the child pointer changes 
between two successive reads.  Thus, there are at most $c$ re-reads  caused by each
change to a child pointer (namely by those \op{Refresh} operations running when the change happens).  Moreover, there is at most one child pointer change for each update operation.
Thus, the amortized running time per update operation remains $O(h+c)$.
Since queries begin by taking a snapshot of the Version tree, 
queries are wait-free and can be accomplished in the same time that they would 
require in the sequential setting.  For example, searches and order-statistic queries
take $O(h)$ steps.  

\subsection{Extensions}

The variants of the trie described in \Cref{variants} apply equally to the BST.

The approach of \Cref{faster-queries}
can  be applied to our BST in exactly the same way so that,
even though the Node tree is unbalanced, \x{Root}.\x{version}
points to a \emph{balanced} Version tree containing the elements of the set.
This facilitates queries that can be done in the same time
as in a sequential augmented balanced BST.  For example,
order-statistic queries can all be answered in $O(\log n)$ time
where $n$ is the size of the set.
This does, however, increase the amortized time for update operations, which can be bounded using
the argument of \Cref{faster-queries} by $O((h+c)\log{\hat n})$, where $\hat{n}$ is a bound on the size of the set
under any possible linearization of the execution.
%\here{explain?}

% !TEX root =  augmented-tree.tex

\section{Future Work}

Our technique  provides lock-free implementations of many tree data structures
based on augmented trees supporting insertions, deletions, and
arbitrarily complex queries.

Although we base our augmented BST on \cite{EFHR14}, we believe our technique could also be applied to the similar
lock-free BST design of Natarajan, Ramachandran and Mittal \cite{NRM20} or other lock-free trees.
It would be interesting to apply it to 
a balanced tree such as the lock-free chromatic BST of \cite{BER14} or to a self-balancing concurrent tree such as the CB Tree \cite{AKK+14}.
In particular, this would require ensuring the \op{Propagate} routine
works correctly with rotations used to rebalance the tree.
The technique may also be applicable to trees that use other coordination mechanisms, such as locks (e.g., \cite{NRM20}).

%\here{Can this approach be used on arbitrary pointer-based data structures to make them snapshottable?  It's a lot simpler than our multiversion work with CMU because we don't need version lists and we don't need to clean out old versions--it happens automagically.}

Could our technique be extended to obtain lock-free implementations of 
sequential augmented data structures that require more complex updates
(such as the insertion of a pair of keys)?
In the sequential setting, examples of such data structures  include
link/cut trees \cite{ST83} and segment trees \cite{Bendl77,Berg08}.
%\here{double check that these are the best examples for this.
%2-d range trees that support queries for the number of points in a given rectangle \here{\cite{?}}.
Shafiei \cite{Sha19} described a mechanism for making multiple changes to a tree 
appear atomic, but it would require additional work to find a suitable way to generalize
our \op{Propagate} routine with her approach.

%\paragraph{Acknowledgements}
%We thank Shahin Kamali
%This research was supported by the Natural Science and Engineering Research Council of Canada
%\here{add Youla's funding sources}

\appendix
% !TEX root =  augmented-tree.tex

\section{Proof of Correctness for the Wait-Free Static Trie}
\label{static-proof}

In this appendix, we give a full proof of linearizability for the augmented trie of \Cref{static}.
Recall that $U_x$ is the sequence of keys from the universe $U$ that are represented in the leaves of the
subtree of Nodes rooted at Node $x$
in the order they appear from left to right.
More formally, we can define $U_x$ inductively from the bottom of the tree to the top.
\begin{eqnarray*}
U_{\textit{Leaf}[k]} &=& \langle k\rangle \hspace*{18mm}\mbox{ if } 0\leq k\leq N-1\\
U_x &=& U_{x.\textit{left}}\cdot U_{x.\textit{right}} \mbox{ if $x$ is an internal Node} 
\end{eqnarray*}
%We use $U_x$ later for referring to the keys represented by the leaves of the trees of Versions 
%rooted at $x.\v{version}$.

We start with a straightforward invariant that, for any Node $x$,
the subtree of Versions stored rooted at $x.\x{version}$
has the same shape as the subtree of Nodes rooted at $x$.

\begin{invariant}\label{shape}
For any Node $x$, the tree of Version objects rooted at $x.\x{version}$ is a perfect binary tree with $|U_x|$ leaves.
\end{invariant}
\begin{proof}
The initialization of our data structure ensures that this is true in the initial configuration.
We argue that every update to the \x{version} field of a $Node$ preserves the invariant.
If a \CAS\ on line \ref{cas-ins} or \ref{cas-del} updates the \x{version} field of a leaf Node,
it installs a pointer to a Version object created on line \ref{new-leaf-ins} or \ref{new-leaf-del} with no children,
so the invariant is preserved.
If a \CAS\ on line \ref{cas-refresh} installs a new Version $v$ in the \x{version} field of an internal Node $x$ at height $h$,
$v$'s left and right subtrees $T_L$ and $T_R$ are read from $x.\x{left}.\x{version}$ and $x.\x{right}.\x{version}$ at
lines \ref{read-left} and \ref{read-right}.  Assuming the claim holds at all times before the \CAS, $T_L$ and $T_R$ are 
perfect binary trees with $|U_{x.\textit{left}}|=2^{h-1}$ and $|U_{x.\textit{right}}|=2^{h-1}$ leaves, respectively, so $v$ is the root of a perfect binary 
tree with $2^h=|U_x|$ leaves.
\end{proof}

The following invariant that $sum$ fields are correctly computed is trivial, since it is satisfied whenever an internal Version is created at 
line  \ref{new-node}, and fields of Versions are immutable.

\begin{invariant}\label{sums-correct}
For every internal Version $v$, $v.\x{sum} = v.\x{left}.\x{sum} + v.\x{right}.\x{sum}$.
\end{invariant}

\subsection{Linearization Respects Real-Time Order}

For $k\in U$, let $Path(k)$ be the path of Nodes in the tree from $\x{Leaf}[k]$ up to the root.
In this section, we show that arrival points of update operations on key $k$
at all Nodes along $Path(k)$ are during the execution interval of the update.
In particular, this means each update's linearization point  is during the update,
since we define the linearization point to be the arrival point at the root Node.
This ensures that if one update ends before another begins, the first appears earlier
in the linearization, as required by the definition of linearizability.

The first easy lemma shows the arrival points are after the invocation of the update.

\begin{lemma}\label{after-start}
The arrival point of an update operation on a key $k$ at a node $x$ in $Path(k)$ is after the start of the update.
\end{lemma}
\begin{proof}
Consider an update operation $U$ that is either an \op{Insert}($k$) or a \op{Delete}($k$) operation.
Let $x_0, \ldots, x_m$ be the Nodes of $Path(k)$.
We prove the claim holds for all $x_i$ by induction on $i$.
For the base case, it follows from \cref{arr-leaf-suc,arr-leaf-fail,arr-leaf-read} of
\Cref{arrival} that $U$'s arrival point at $x_0=\x{Leaf}[k]$ is after the start of $U$.
For the induction step, if the claim holds for $x_{i-1}$, it follows from 
\cref{arr-left,arr-right} of \Cref{arrival} that it holds for $x_i$,
since $U$ must arrive in $x_{i-1}$ before it can arrive in $x_i$.
\end{proof}

The next few results lead to \Cref{before-end}, which states that every update operation has an arrival point at the root Node before it terminates.
The following observation is an immediate consequence of \cref{arr-left,arr-right} of \Cref{arrival}.
\begin{observation}\label{succ-refresh}
For any call to \op{Refresh}($x$) on an internal Node $x$ that performs a successful \CAS\ at line \ref{cas-refresh},
each operation whose arrival point at a child of $x$ is before the \op{Refresh}$(x)$ performs line \ref{read-old}
has an arrival point at $x$ no later than the \CAS\ of that \op{Refresh}.
\end{observation}

\Cref{arrival} ensures that the arrival point of an update operation at a Node is unique if it exists.  
The following
Lemma shows that the arrival point of a completed update on key $k$ exists at all Nodes along $Path(k)$.
To prove this, we observe that an update operation with key $k$ calls \op{Propagate}, which performs a double \op{Refresh} on 
all the ancestors of $\x{Leaf}[k]$.
We show in \Cref{before-end} that the update has an arrival point at each ancestor before the end of the double \op{Refresh} on it.  It is proved using the following simple observation.

\begin{observation}\label{non-overlapping}
If two calls to \op{Refresh} on the same Node both perform successful \op{CAS} steps,
then one performs the read on line \ref{read-old} after the \op{CAS} on line \ref{cas-refresh} of the other.
\end{observation}
\begin{proof}
Let $x$ be an internal Node.
Only a successful \op{CAS} on line \ref{cas-refresh} can update $x.\x{version}$.
Whenever this happens, a pointer to a Version object that was newly created on line \ref{new-node} is stored in $x.\x{version}$.
Thus, the value stored in $x.\x{version}$ was never stored there before.
It follows that there cannot be a successful \op{CAS} between the time
$x.\x{version}$ is read on line \ref{read-old} of a \op{Refresh}($x$) 
and the time that \op{Refresh} performs a successful \op{CAS} on line \ref{cas-refresh}.
The claim follows.
\end{proof}

\begin{lemma}\label{before-end}
Each completed update operation has an arrival point at the root Node before the end of the update.
\end{lemma}
\begin{proof}
Consider an \op{Insert}($k$) or \op{Delete}($k$) operation $U$.
Let $x_0, \ldots, x_m$ be $Path(k)$.

We show by induction that for $0\leq i\leq m$, the update has an arrival point at $x_i$ before $U$'s invocation of \op{Propagate} has completed
$i$ iterations of the loop in lines \ref{prop-loop-start}--\ref{prop-loop-end}.

For the base case, it follows from \cref{arr-leaf-suc,arr-leaf-fail,arr-leaf-read} of
\Cref{arrival} that $U$'s arrival point at $\x{Leaf}[k]$ is before the time $U$
calls \op{Propagate} on line \ref{prop-ins} or \ref{prop-del}.
So, the claim holds after 0 iterations of the loop.

For the induction step,
let $i>0$ and assume the claim holds after $i-1$ iterations of the loop.
The $i$th iteration of the loop does a double \op{Refresh} on $x_i$.
If either invocation of \op{Refresh}($x_i$)  at line \ref{refresh-1} or \ref{refresh-2}  
of the $i$th loop iteration
performs a successful \CAS\ then the claim follows from \Cref{succ-refresh}.
Otherwise, let $R_1$ and $R_2$ be the two invocations of \op{Refresh}($x_i$), which both 
perform an unsuccessful \CAS.
Then, there must have been two successful \CAS\ steps $c_1$ and $c_2$ by other processes
between line \ref{read-old} and \ref{cas-refresh} of $R_1$ and $R_2$, respectively.
(See \Cref{before-end-fig}.)
By \Cref{non-overlapping}, the invocation $R$ of \op{Refresh}($x_i$) that 
performed $c_2$ must have read $x_i.\x{version}$ at line \ref{read-old}
\emph{after} $c_1$, which is after the beginning of $R_1$.
By the induction hypothesis, $U$ has an arrival point at $x_{i-1}$ before the beginning of $R_1$
and therefore before $R$ performs line \ref{read-old}.
Thus, by \Cref{succ-refresh} applied to $R$, $U$ has an arrival point at $x_i$ 
that is no later than $R$'s \op{CAS}, which is before the end of $R_2$.
The claim follows for $x_i$.
\end{proof}

\Cref{before-end} shows that each completed update has 
a well-defined linearization point before it terminates.
Together with \Cref{after-start}, this means that the linearization point of each
update is during its execution interval. 

\subsection{Linearization is Consistent with Responses}

The goal of this section is to prove that operations in the concurrent execution 
return the same results that they would if they were performed sequentially
in their linearization order.
To do this, we prove \Cref{inv:correct}, which says that for every Node $x$, the Version tree
rooted at $x.\x{version}$ stores the set of keys that would be present
after performing the operations with arrival points at $x$, in the order of their
arrival points.

The following Lemma is helpful for proving \Cref{inv:correct} for leaf Nodes.
It shows that groups of operations that arrive at a leaf at the same time are of the same type.

\begin{lemma}\label{sametype}
Let $k$ be any key.
If the arrival points of several operations at the leaf Node $x$ for key $k$ are at the same step, then the operations are either all \op{Insert}($k$) operations or all \op{Delete}($k$) operations.
\end{lemma}
\begin{proof}
Multiple operations can be assigned the same arrival points at the leaf Node $x$ only at a successful \CAS\ step on $x.\x{version}$.
Consider a successful \CAS\ $s$ performed at line \ref{cas-ins} of an \op{Insert}($k$) operation $I$.
(The argument if the \CAS\ is performed by an \op{Delete} is similar.)
Then, \cref{arr-leaf-suc,arr-leaf-fail} of \Cref{arrival}
specify that $s$ is the arrival point at $x$ of $I$ and some update operations 
that fail their \CAS\ on $x.\x{version}$.
Since $s$ succeeds, the value of $x.\x{version}$ in the configuration $C$ before $s$ is the same value that $I$ read on line \ref{read-old-ins}.
Since $I$'s test on line \ref{test-ins} was true, $x.\x{version}.\x{sum}$ must have been 0 in $C$.
Let $U$ be another update operation whose arrival point at $x$ is $s$.
We argue that $U$ must be an \op{Insert}($k$).
By \cref{arr-leaf-fail} of \Cref{arrival}, $s$ is the first successful \CAS\ on $x.\x{version}$ after $U$ reads $x.\x{version}$ on line \ref{read-old-ins} or \ref{read-old-del}.
Thus, $U$ reads the same Version from $x.\x{version}$ as is still there in $C$.
If $U$ were a \op{Delete} operation, the test on line \ref{test-del} would return false 
and $U$ would not perform its \CAS\ on line \ref{cas-del}.
\end{proof}

For any configuration $C$, key $k$ and Node $x$ in $Path(k)$,
let $Ops(C,x,k)$ be the sequence of update operations of the form \op{Insert}($k$) or \op{Delete}($k$) with arrival points at  $x$ prior to $C$,
in the order of their arrival points at $x$.

%\here{draw pictures of an example to illustrate $Ops$ to explain what happens when some operations are promoted from child to parent.}

The following lemma is useful for showing that updates on a particular key
are linearized in the same order that they arrive at the leaf for that key,
which will allow us to show that the results returned
by operations are consistent with their linearization order.

\begin{invariant}\label{parent-same}
Let $C$ be any configuration.
If Node $x$ is a non-root ancestor of the leaf Node for key $k$, then $Ops(C,x.\x{parent},k)$ is a prefix of $Ops(C,x,k)$.
\end{invariant}
\begin{proof}
In the initial configuration $C_0$, $Ops(C_0,x,k)$ and $Ops(C_0,x.\x{parent},k)$ are empty sequences, so the claim holds.
Assuming the invariant holds at some configuration $C$, we show that it holds at the next configuration $C'$.
If a step extends the sequence of operations with arrival points at $x$, the invariant is clearly preserved.
Consider a step that extends the sequence of operations with arrival points at $x.\x{parent}$.
Since $Ops(C,x.\x{parent},k)$ is a prefix of $Ops(C,x,k)$ and \Cref{arr-left,arr-right} of \Cref{arrival}
imply that $Ops(C',x.\x{parent},k)$ is obtained
by appending to $Ops(C,x.\x{parent},k)$ all operations of $Ops(C,x,k)$ that do not already
appear in $Ops(C,x.\x{parent},k)$ (in the order they
appear in $Ops(C,x,k)$), we have $Ops(C',x.\x{parent},k)=Ops(C,x,k)=Ops(C',x,k)$ so the invariant is satisfied in $C'$.
\end{proof}

%\here{Check:  since there is a bit of asymmetry between insert(k) and delete(k) (because key k is initially not in the data structure) check that whenever we said delete is similar to insert that there is nothing to add to take care of this.}

\begin{corollary}\label{root-same}
Let $C$ be any configuration.
If $x$ is the leaf Node for key $k$, then $Ops(C,\x{Root},k)$ is a prefix of $Ops(C,x,k)$.
\end{corollary}
\begin{proof}
This follows from applying \Cref{parent-same} repeatedly along the path from $x$ to the root Node.
\end{proof}

Now we are ready to prova a key invariant for correctness.
\Cref{shape} ensures that the leaves referred to in \Cref{inv:correct}, below, are well-defined,
and that \Cref{inv:correct} fully specifies what values the \x{sum} fields of all of the leaves
in Version trees have.

\begin{invariant}\label{inv:correct}
The following holds in every configuration $C$.
For each Node $x$, and $1\leq i\leq |U_x|$, if $k$ is the $i$th $key$ in $U_x$
then the $i$th leaf in the tree of Version objects rooted at $x.\x{version}$ has \x{sum} 1 if and only if
 $Ops(C,x,k)$ is non-empty and its last operation is an \op{Insert}($k$).
\end{invariant}
\begin{proof}
In the initial configuration $C_0$, $Ops(C_0,x,k)$ is empty and all Versions are initialized with \x{sum} field equal to 0.

Assume the claim holds in all configurations before a step $s$ to prove that it holds in 
the configuration after $s$.
Let $C$ be the configuration immediately before $s$ and $C'$ be the configuration immediately after $s$.
The  truth of the invariant can be affected by $s$ only if $s$ is a successful \CAS\ step on the \x{version} field of a Node
or the arrival point of some operation(s) at a Node.
We consider all such steps in several cases.

Suppose $s$ is a read step at line \ref{read-old-ins} that is the arrival point of an \op{Insert}($k$) at $k$'s leaf Node $x$,
in accordance with \cref{arr-leaf-read} of \Cref{arrival}.
Then, $x.\x{version}.\x{sum} = 1$ in both $C$ and $C'$.
Because the invariant holds in $C$, 
and $Ops(C',x,k) = Ops(C,x,k)\cdot \op{Insert}(k)$, the invariant holds in $C'$ at $x$.

The argument if $s$ is a read step at line \ref{read-old-del} that is the arrival point of a \op{Delete}($k$) at $k$'s leaf is symmetric.

Suppose $s$ is a successful \CAS\ on the \x{version} field of a leaf Node $x$ at line \ref{cas-ins}.
By \Cref{sametype}, the operations whose arrival points are at $s$ are all \op{Insert}($k$) operations.
Thus, $Ops(C',x,k)$ is obtained from $Ops(C,x,k)$ by appending one or more \op{Insert}$(k)$ operations to the end of the sequence.
Moreover, $x.\x{version}.\x{sum}=1$ in $C'$, so the claim holds for $x$ in $C'$.

The argument if $s$ is a successful \CAS\ on the \x{version} field of a leaf $x$ at line \ref{cas-del} is symmetric.

Now, suppose $s$ is a successful \CAS\ of some call $R$ of \op{Refresh}($x$)
that stores the Version $v$ in the \x{version} field of an internal Node $x$.
According to \cref{arr-left,arr-right} of \Cref{arrival},
this \CAS\ may be the arrival point in $x$ of some number of updates involving keys in $U_x$.

We first consider the case where $k$ is in $U_{x.\textit{left}}$.
Prior to $R$'s \CAS, $R$ reads some Version $v_L$ from $x.\x{left}.\x{version}$ on line \ref{read-left}.
Let $C_L$ be the configuration immediately after this read.
Line \ref{new-node} sets $v.\x{left} = v_L$.
It follows from \Cref{non-overlapping}
that any call to \op{Refresh}($x$) that performs a successful
\CAS\ on $x.\x{version}$ strictly before $s$ 
does its \op{CAS}, and that \CAS\ is before $C_L$.
Hence, any such \op{Refresh} also reads $x.\x{left}.\x{version}$ on line \ref{read-left}
before $C_L$.
It follows from \cref{arr-left} of \Cref{arrival} and \Cref{parent-same} that
$Ops(C',x,k) = Ops(C_L,x.\x{left},k)$.
Moreover, the $i$th key of $U_x$ is also the $i$th key of $U_{x.\textit{left}}$
and the $i$th leaf of the tree rooted $v$ is the $i$th leaf of the
tree rooted at $v_L$.
So, the last operation in $Ops(C',x,k)$ is an \op{Insert}($k$) if and only if
the last operation in $Ops(C_L,x.\x{left},k)$ is an \op{Insert}($k$), by our assumption that the claim
holds in configuration $C_L$ before $s$.
The latter statement is true if and only
if the \x{sum} field of the $i$th leaf of the subtree rooted at $v_L$ is 1,
which is true if and only if the \x{sum} field of the $i$th leaf of the subtree rooted at $v$ is 1.
This proves that the claim holds at $x$ in $C'$ because $s$ stores $v$ in $x.\x{version}$.

The argument for the $(i-|U_{x.\textit{left}}|)$th key in $U_{x.\textit{right}}$, which is the $i$th key in $U_x$, is similar.
\end{proof}

\begin{lemma}\label{updates-correct}
Each completed update operation returns the same result as it would if all updates were performed sequentially in their linearization order.
\end{lemma}
\begin{proof}
Consider a completed \op{Insert}($k$) operation $I$.  (The argument for a \op{Delete} operation is similar.)
Let $C$ be the configuration after $I$'s arrival point at the root Node and let $x$ be the leaf Node corresponding to key $k$.
$I$ should return false if and only if the last update operation on $k$ preceding it in the linearization order is an \op{Insert}($k$).
Since operations are linearized according to their arrival points at the root Node,
this is equivalent to saying that the last operation before $I$ in $Ops(C,\x{Root},k)$ is an \op{Insert}.
By \Cref{root-same}, this is equivalent to saying that the last operation before $I$ in $Ops(C,x,k)$ is an \op{Insert}.
So, we must show that $I$ returns false if and only if the last operation before $I$ in $Ops(C,x,k)$ is an \op{Insert}.

First, suppose $I$ reads a Version whose \x{sum} is 1 at line \ref{read-old-ins}.
Then, $I$ returns false.
Moreover, in the configuration $C_1$ before $I$'s arrival point at $x$, $x.\x{version}.\x{sum}=1$, so by \Cref{inv:correct}
the last operation in $Ops(C_1,x,k)$ is an \op{Insert}($k$).
Thus, since $I$ is the only operation whose arrival point at $x$ is in the step after $C_1$ (by \cref{arr-leaf-read} of \Cref{arrival}),
the last operation in $Ops(C,x,k)$ before $I$ is an \op{Insert}($k$), as required.

For the remainder of the proof, suppose $I$ reads a Version whose \x{sum} is 0 at line \ref{read-old-ins}.

Now, suppose $I$ performs an unsuccessful \CAS\ on line \ref{cas-ins}.  
In this case, $I$ also returns false.  According to \Cref{arrival},
$I$ is among several updates whose arrival points are at one successful \CAS, and $I$ is not the first in this group.
By \Cref{sametype}, they are all \op{Insert}$(k)$ operations, so the last operation before $I$ in
$Ops(C,x,k)$ is an \op{Insert}, as required.

Finally, suppose $I$ performs a successful \CAS\ on line \ref{cas-ins}.
In this case, $I$ returns \true.
Let $C_2$ be the configuration before that \CAS.
Since the \CAS\ succeeds, the value of $x.\x{version}$ in $C$ is the same as it was when $I$ read 
it at line \ref{read-old-ins} and $C$.
So, in $C$, $x.\x{version}.\x{sum}=0$.
Thus, $Ops(C_2,x,k)$ is either empty or ends with a \op{Delete} operation.
Since $I$ is the first operation whose arrival point at $x$ is at the successful \CAS,
the last operation before $I$ in $Ops(C,x,k)$ is not an \op{Insert}, as required.
\end{proof}

It remains to show that query operations can be linearized.
We assume that a query begins by reading a Version $v$ from $\x{Root}.\x{version}$, and then performs the same code
on the tree of Versions rooted at $v$ as it would in the sequential data structure.
The linearization point of the query is when it reads $\x{Root}.\x{version}$.
It follows from \Cref{inv:correct} that the leaves of the tree rooted at $v$ accurately reflect the
contents of the represented set $S$ at the query's linearization point.  
By \Cref{sums-correct}, all Versions in the Version tree rooted at $v$ have \x{sum} fields
that accurately reflect the \x{sum} fields of the leaves, so queries on this
tree will return results consistent with their linearization points.

%We must also ensure that once an operation is propagated to some Node $x$, that subsequent
%\CAS\ instructions on $x.\x{version}$ do not incorrectly erase the effect of the operation.
%For the latter goal, one key observation is that if two CAS instructions update $x.\x{version}$ then
%the \op{Refresh} that performs the second \CAS\ reads $x.\x{left}.\x{version}$ and $x.\x{right}.\x{version}$ later than
%the \op{Refresh} that performs the first \CAS.  
%An inductive argument then shows that because operations do not get erased at one level of the tree,
%they do not get erased at the next higher level.

\section{Pseudocode for Faster Queries in Wait-free Trie}
\label{faster-queries-app}

Here, we provide some more detailed information about the modification outlined in \Cref{faster-queries} to 
improve the time for searches and other order statistic
queries to $O(\log |S|)$.
These modifications apply equally to the trie of \Cref{static} and the BST of \Cref{bst-sec}.
The type definitions are given in \Cref{faster-queries-code}.
The only difference for the Node type is that the $version$ field now stores a pointer to (the root of) a RBT.
We remark that the RBT is a standard node-oriented RBT where keys are stored in internal nodes as well as leaves.
RBT nodes are represented using the RNode type.
Each RBT node is augmented with a $sum$ field that stores the number of elements in the subtree rooted at
that node to facilitate order-statistic queries.
All fields of an RBT node are immutable.

The initialization is the same as in \Cref{pseudocode}, except that all Node's $version$ field
can initially point to a single dummy Rnode with $sum$ field 0 and \nil\ child pointers.
The \op{Insert} and \op{Delete} routines are identical to \Cref{pseudocode},
except that lines \ref{new-leaf-ins} and \ref{new-leaf-del} create a new RNode instead of a Version
object.  Line \ref{new-leaf-ins} must fill in $new$'s additional fields $key\leftarrow k$ and 
$colour \leftarrow \mbox{black}$.
The modified \op{Refresh} routine is shown in \Cref{faster-queries-code}.
It uses a non-destructive implementation
of the standard (sequential)  \op{Join} algorithm to combine two RBTs into one.
The \op{Propagate} routine is identical to \Cref{pseudocode}.
Queries are done by reading $\x{Root}.\x{version}$ and running standard (sequential) RBT query algorithms
on it.
%\here{figure out how to shade changes to code later}

\begin{figure}[t]
\begin{algorithmic}[1]
\StartFromLine{200}
\Type Node \Comment{used to store nodes of static trie structure}
	\State Node* \x{left}, \x{right} \Comment{immutable pointers to children Nodes}
	\State Node* \x{parent}		\Comment{immutable pointer to parent Node}
	\State Rnode* \x{version}  \Comment{mutable pointer to RBT}	
\EndType

\medskip

\Type Rnode \Comment{used to store a node of an RBT; replaces Version objects}
	\State Rnode* \x{left}, \x{right} \Comment{immutable pointers to children}
	\State int \x{sum} \Comment{immutable number of keys in subtree rooted at Rnode}
	\State U \x{key} \Comment{immutable key needed for searching RBT}
	\State $\{\mbox{red, black}\}$ \x{colour} \Comment{immutable colour used for balancing RBT}
\EndType

\medskip

\Function{Boolean}{Refresh}{Node* $x$} \Comment{Try to propagate information to Node $x$ from its children}
	\State $\x{old} \leftarrow x.\x{version}$
	\State $v_L \leftarrow x.\x{left}.\x{version}$
	\State $v_R \leftarrow x.\x{right}.\x{version}$
	\If{$v_L.sum = 0$} $new \leftarrow v_R$
	\ElsIf{$v_R.sum = 0$} $new \leftarrow v_L$
	\Else\ $new \leftarrow \op{Join}(v_L,v_R)$ \Comment{Non-destructive join of two RBTs into new RBT}
	\EndIf
	\State \Return $\CAS(x.\x{version}, \x{old}, \x{new})$
\EndFunction

\end{algorithmic}
\caption{Modified type definitions and \op{Refresh} routine to support faster queries.\label{faster-queries-code}}
\end{figure}

% !TEX root =  augmented-tree.tex

\section{Pseudocode for Lock-Free Augmented BST}
\label{BST-pseudo-section}

%\here{Should we try to split description of algorithms into original BST and augmentation?}
%\here{should there be something about uniqueness of arrival points?}

Here, we give more details about how to augment the lock-free BST of Ellen et al.~\cite{EFHR14}.
Type definitions are given in \Cref{bst-types}.
High-level pseudocode for \op{Insert} and \op{Delete} are given in \Cref{BST-pseudo-1}.
These are mostly the same as in \cite{EFHR14}, except for the addition of calls to \op{Propagate}
and the creation of Version objects used to initialize the \x{version} fields of Nodes created
by \op{Insert}.
Consequently, we do not give all the details of these routines; see \op{EFHR14} for the detailed pseudocode.
The new routines for handling Versions and some example queries are given in \Cref{BST-pseudo-2}.

An \boldop{Insert}($k$) searches for $k$ in the BST of Nodes and arrives at a leaf Node $\ell$ containing some key $k'$.
If $k'=k$, the value $k$ is already in the BST, so the \op{Insert} does not need to modify the tree and will eventually return false.
Otherwise, the \op{Insert} attempts to replace the leaf $\ell$ by a new internal Node
whose key is $\max(k,k')$ with two new leaf children whose keys are $\min(k,k')$ and $\max(k,k')$.
There are also some additional steps required to coordinate updates to the same part of the tree, and those steps
may cause the attempt to fail, in which case the \op{Insert} tries again by backtracking
up the tree and then searching down the tree for the correct place to try inserting the node again.
The details of the inter-process coordination are not important to the augmentation.
Before attempting to add the three new Nodes to the tree, the \op{Insert} creates a new Version object 
for each of them with fields filled in as shown in \Cref{succ-ins-fig}.
To facilitate backtracking after an unsuccessful attempt, the \op{Insert}
keeps track of the sequence of internal Nodes visited on the way
to the location to perform the insertion in a \emph{thread-local} stack.
When an attempt of the \op{Insert} succeeds, it calls \op{Propagate} on the newly inserted internal Node and returns \true.
\op{Propagate} uses the thread's local stack to revisit the Nodes along the path from
the root to the insertion location in reverse order, performing a double \op{Refresh} on each Node, as in \Cref{static}.
If the \op{Insert} terminates after finding the key is already present in a leaf Node, 
it calls \op{Propagate} on that leaf Node, to ensure that the operation that inserted 
that leaf Node has been linearized, and then returns false.

\begin{figure}[t]
\begin{algorithmic}[1]
\StartFromLine{100}
\Type Node \Comment{used to store nodes of static trie structure}
	\State $U$ \x{key} \Comment{immutable key of Node}
	\State Node* \x{left}, \x{right} \Comment{mutable pointers to children Nodes}
	\State Version* \x{version}  \Comment{mutable pointer to current Version}	
	\State Info* \x{info} \Comment{for coordinating updates; irrelevant to our augmentation}
\EndType

\medskip

\Type Version \Comment{used to store a Node's augmented data}
	\State $U$ \x{key} \Comment{immutable key of Node this Version belongs to}
	\State Version* \x{left}, \x{right} \Comment{immutable pointers to children Versions}
	\State int \x{sum} \Comment{immutable sum of descendant leaves' bits}
\EndType
\end{algorithmic}
\caption{Object types used in lock-free augmented BST data structure.\label{bst-types}}
\end{figure}

\begin{figure}
\begin{algorithmic}[1]
\StartFromLine{109}
\State Initialize the data structure as shown in \Cref{bst-init}, where $Root$ is a shared pointer\label{bst-initialize}

\medskip

\Function{Boolean}{Insert}{Key $k$}
	\State let $stack$ be an empty thread-local stack
	\State push \x{Root} on to \x{stack}
	\Loop
		\State do a BST search for $k$ from top Node on \x{stack}, pushing visited internal Nodes on~\x{stack}
		\State let $\ell$ be the leaf reached by the search 
		\If{$\ell.\x{key} = k$} \label{ins-test}
			\State \newcode{\textsf{Propagate}(\textit{stack})}\label{ins-prop-1}
			\State \Return false \Comment{$k$ is already in the tree}\label{ins-return-false}
		\EndIf
		\State let $p$ be the top Node $p$ on \x{stack} \Comment{$p$ was $\ell$'s parent during the search}
		\State let $\x{new}$ be a new internal Node whose children are a new leaf Node with \x{key} $k$ and a\label{new-triangle}
			\State \hspace*{4mm}   new Leaf with $\ell$'s \x{key}.  
				   \newcode{Each of the three new Nodes has a pointer to a new Version}
			\State \hspace*{4mm} \newcode{object with the same \textit{key} as the Node. The leaf Versions have \textit{sum} 1 (or 0 if the key}
			\State \hspace*{4mm} \newcode{is $\infty_1$ or $\infty_2$) and $\textit{new}.\textit{sum}=\textit{new}.\textit{left}.\textit{sum}+\textit{new}.\textit{right}.\textit{sum}$.} (See \Cref{succ-ins-fig}.)
		\State attempt to change $p$'s child from $\ell$ to \x{new} using \op{CAS}
		\If{attempt fails due to another update} 
			\State help complete the update that caused the attempt to fail
			\State backtrack by popping \x{stack} until a node that is not marked for deletion is popped, 
				\State \hspace*{4mm}helping complete the deletion of each marked Node that is popped
		\Else	\Comment{\x{new} was successfully added to tree}
			\State \newcode{\textsf{Propagate}(\textit{stack})}\label{ins-prop-2}
			\State \Return \true\label{ins-return-true}
		\EndIf
	\EndLoop
\EndFunction

\medskip

\Function{Boolean}{Delete}{Key $k$}
	\State let $stack$ be an empty thread-local stack
	\State push $\x{Root}$ on to \x{stack}
	\Loop
		\State do a BST search for $k$ from top Node on \x{stack}, pushing visited internal Nodes on~\x{stack}
		\State let $\ell$ be the leaf reached by the search 
		\If{$\ell.\x{key} \neq k$} \label{del-test}
			\State \newcode{\textsf{Propagate}(\textit{stack})}\label{del-prop-1}
			\State \Return false \Comment{$k$ is not in the tree}\label{del-return-false}
		\EndIf
		\State pop Node $p$ from \x{stack} \Comment{$p$ was $\ell$'s parent during the search}
		\State let $gp$ be the top Node on \x{stack} \Comment{$gp$ was $p$'s parent during the search}
		\State attempt to change $gp$'s child from $p$ to $\ell$'s sibling using \op{CAS}
		\If{attempt fails due to another update} 
			\State help complete the update that caused the attempt to fail
			\State backtrack by popping \x{stack} until a node that is not marked for deletion is popped, 
				\State \hspace*{4mm}helping complete the deletion of each marked Node that is popped
		\Else	\Comment{deletion removed $k$'s Node from tree}
			\State \newcode{\textsf{Propagate}(\textit{stack})}\label{del-prop-2}
			\State \Return \true\label{del-return-true}
		\EndIf
	\EndLoop
\EndFunction

\end{algorithmic}
\caption{Pseudocode for augmented BST.  The code for updates is given at a high level.  For details, see \cite{EFHR14}.  Changes to \op{Insert} and \op{Delete} to support augmentation is \newcode{shaded}.\label{BST-pseudo-1}}
\end{figure}

\begin{figure}
\begin{algorithmic}[1]
\StartFromLine{152}
%\State \textbf{Simple version:}
%\Function{Boolean}{Refresh}{Node* $x$} \Comment{Try to propagate information to Node $x$ from its children}
%	\State $\x{old} \leftarrow x.\x{version}$\label{read-old'}
%	\State $x_L \leftarrow x.\x{left}$\label{read-left-child}
%	\State $v_L \leftarrow x_L.\x{version}$\label{read-left-version}
%	\State $x_R \leftarrow x.\x{right}$\label{read-right-child}
%	\State $v_R \leftarrow x_R.\x{version}$\label{read-right-version}
%	\State $\x{new} \leftarrow$ new Version with $\x{left}=v_L$, $\x{right}=v_R$, $\x{sum}=v_L.\x{sum}+v_R.\x{sum}$\label{new-node'}
%	\State \Return $\CAS(x.\x{version}, \x{old}, \x{new})$\label{cas-refresh'}
%\EndFunction
%
%\medskip
%
%\State \textbf{Version with repeated reads:}
\Function{Boolean}{Refresh}{Node* $x$} \Comment{Try to propagate information to Node $x$ from its children}
	\State $\x{old} \leftarrow x.\x{version}$\label{read-old'}
	\Repeat \Comment{Get a consistent view of $x.\x{left}$ and $x.\x{left}.\x{version}$}
		\State $x_L \leftarrow x.\x{left}$\label{read-left-child}
		\State $v_L \leftarrow x_L.\x{version}$\label{read-left-version}
	\Until{$x.\x{left} = x_L$}
	\Repeat \Comment{Get a consistent view of $x.\x{right}$ and $x.\x{right}.\x{version}$}
		\State $x_R \leftarrow x.\x{right}$\label{read-right-child}
		\State $v_R \leftarrow x_R.\x{version}$\label{read-right-version}
	\Until{$x.\x{right} = x_R$}
	\State $\x{new} \leftarrow$ new Version with $\x{left}\leftarrow v_L$, $\x{right}\leftarrow v_R$, $\x{sum}\leftarrow v_L.\x{sum}+v_R.\x{sum}$\label{new-node'}
	\State \Return $\CAS(x.\x{version}, \x{old}, \x{new})$\label{cas-refresh'}
\EndFunction

\medskip

\Function{}{Propagate}{Stack* \x{stack}} \Comment{Propagate updates starting at top Node on \x{stack}}
	\While{\x{stack} is not empty}\label{prop-loop-start'}
		\State pop Node $x$ off of \x{stack} \label{go-up'}
		\If{not $\op{Refresh}(x)$}\label{refresh-1'}
			\State $\op{Refresh}(x)$\Comment{Do a second \op{Refresh} if first one fails}\label{refresh-2'}
		\EndIf
	\EndWhile\label{prop-loop-end'}
\EndFunction
\medskip

\Function{Boolean}{Find}{$k$} \Comment{Returns \true\ if $k$ is in the set, or false otherwise}
	\State $v\leftarrow \x{Root}.\x{version}$
	\While{$v.\x{left}\neq\nil$}
		\If{$k<v.\x{key}$} $v\leftarrow v.\x{left}$
		\Else\ $v\leftarrow v.\x{right}$
		\EndIf
	\EndWhile
	\State \Return ($v.\x{key} = k$)
\EndFunction

\medskip

\Function{$U$}{Select}{$j$} \Comment{Returns set's $j$th smallest element or \nil\ if no such element}
	\State $v \leftarrow \x{Root}.\x{version}$
	\If{$j>v.\x{sum}$} \Comment{Set contains fewer than $j$ elements}
		\State \Return \nil 
	\EndIf
	\Repeat \Comment{Loop invariant:  desired element is $j$th in tree rooted at $v$}
		\If{$j\leq v.\x{left}.\x{sum}$}
			\State $v \leftarrow v.\x{left}$
		\Else
			\State $j\leftarrow j-v.\x{left}.\x{sum}$
			\State $v \leftarrow v.\x{right}$
		\EndIf
	\Until{$v$ is a leaf}
	\State \Return $v.\x{key}$
\EndFunction

\medskip

\Function{int}{Size}{}
	\State \Return $\x{Root}.\x{version}.\x{sum}$
\EndFunction
\end{algorithmic}
\caption{Pseudocode augmented BST, continued.  We include \op{Find}, \op{Select} and \op{Size} as three examples of queries that
use the augmentation.\label{BST-pseudo-2}}
\end{figure}

A \boldop{Delete}($k$) has a very similar structure.
It first searches for $k$ in the BST of Nodes and arrives at a leaf Node $\ell$.
If $\ell$ does not contain $k$, then the \op{Delete} does not need to modify the tree and  returns false after calling \op{Propagate}.
Otherwise, the \op{Delete} uses a \CAS\ to attempt to remove both $\ell$ and its parent from the tree.
(See \Cref{succ-del-fig}.)
Again, there are some additional steps required to coordinate updates to the same part of the tree, which may cause
the \op{Delete}'s attempt to fail and retry, but the details are irrelevant to the augmentation.
When an attempt of the \op{Delete} succeeds, it calls \op{Propagate} to perform a double refresh along a path to the root, starting from the internal Node whose child pointer
is changed (i.e., the Node that was formerly the grandparent of the deleted leaf $\ell$) and returns \true.

The \boldop{Refresh}($x$) routine is similar
to the one in \Cref{pseudocode}.
Because the structure of the BST's Node tree can change,
the repeat loops ensure that the \op{Refresh} gets a consistent view of 
$x$'s child pointer and the contents of that child's \x{version} field.
The other difference is that line \ref{new-node'} stores $x.\x{key}$ in the \x{key} field 
of the new Version.  The \boldop{Propagate} routine is identical to the one
given in \Cref{pseudocode}, except that we cannot use parent pointers on line \ref{go-up'}.
Instead, an update operation keeps track of the sequence of Nodes that it traversed from the root
to reach a node $x$ and then does a double \op{Refresh} on each of them in reverse order (from $x$ to the root).

A query operation is performed on a snapshot of the Version tree obtained by reading $\x{Root}.\x{version}$.
This includes the \boldop{Find} operation, which simply
performs a search on the Version tree as it would in a sequential BST.
As an additional bonus, our \op{Find} operation is wait-free,
unlike the original lock-free BST \cite{EFHR14}, where \op{Find} operations may starve.  

\section{Correctness of Augmented BST}
\label{bst-correct-full}

We now give a detailed proof of correctness for the augmented BST of \Cref{bst-sec}.
Throughout this section we consider an execution $\alpha$ of the implementation
and show that it is linearizable.

\subsection{Facts About the Unaugmented BST}

We first summarize some facts from \cite{EFHR14} about the original, unaugmented, lock-free
BST that will be useful for our proof.  
Since our augmentation does not affect the Node tree, these facts remain true in the 
augmented BST.

In \cite{EFHR14}, the mechanism to coordinate updates ensures that a status field
of an internal Node is \emph{flagged} whenever one of the Node's child pointers is changed
and that the Node's status field is permanently flagged before the Node is removed from the tree.  (In \cite{EFHR14}, permanent flags are called ``marks''.)
The following result  is a consequence of this fact.
\begin{lemma}[\changeUnflagged\ of \cite{EFHR14}]\label{change-unmarked}
A Node's child pointer can change only when the Node is not permanently flagged for deletion
and it is reachable.
\end{lemma}

Recall that $T_C$ is the Node tree in configuration $C$ and the search path for a key $k$ in $C$
is the path that would be taken through $T_C$ by a sequential BST search for $k$ in $T_C$.
We say a Node is reachable in $C$ if it appears in $T_C$.

The modifications that can occur in the Node tree are shown in \Cref{tree-changes}.
Neither type of modification can ever give a Node a new ancestor (although a deletion may remove an ancestor).
The next lemma is a consequence of this.

\begin{lemma}[\stillonpath\ of \cite{EFHR14}]\label{still-on-path}
If a Node is on the search path for key $k$ in one configuration and is still reachable in some later configuration, then it is still on the search path for $k$ in the later configuration.
\end{lemma}

When an update operation on key $k$ searches for the location to modify in the tree, it ignores flags.
This means that it may pass through Nodes that have been deleted by concurrent operations.
The following lemma shows that even if this happens, each visited Node \emph{was} on the search 
path for $k$ at some time during the search.  This is crucial for the correctness of the
unaugmented BST, and it is also useful for proving that our augmented tree can be linearized.

\begin{lemma}[\wasOnPath\ of \cite{EFHR14}]\label{was-on-path}
If an \op{Insert}($k$) or \op{Delete}($k$) visits a Node $x$ during its search
for the location of key $k$, then
there was a configuration between the beginning of the operation
and the time it reaches $x$ when $x$ was on the search path for $k$ in the Node tree.
\end{lemma}

An important fact is that all updates to the Node tree preserve the BST property.

\begin{lemma}[\bstProperty\ of \cite{EFHR14}] \label{basic-bst}
In all configurations $C$, $T_C$ satisfies the BST property.
\end{lemma}

\subsection{Linearization Respects Real-Time Order}
\label{bst-order-proof}

In this section, we show that \op{Propagate} succeeds in assigning arrival points to
an update operation at each reachable Node it calls a double \op{Refresh} on,
and that arrival point is during the execution interval of the operation.
In particular, if the call to \op{Propagate} completes, the update is
assigned an arrival point at the root Node that is during the operation.
Since this is used as the linearization point of the update, 
and each query is also assigned a linearization points during the query,
it follows that the linearization respects the real-time order of operations.

%\here{Why is this here}
%Backtracking after an unsuccessful attempt pops Nodes from the thread-local stack to back up to a Node
%that is not marked (and therefore still in the tree) and then searches forward from there.

We first formalize the definition of arrival points described in \Cref{bst-correct-short}.

\begin{definition}
\label{arrival-bst}
We first define the arrival point of each unsuccessful update at a leaf Node.
\begin{enumerate}
\item\label[part]{a-fail}
Consider an \op{Insert}($k$) that arrives at a leaf Node $\ell$ with key $k$ or a \op{Delete}($k$) that arrives at a leaf Node $\ell$ that does not contain $k$.
When searching for the location of $k$, the update arrives at $\ell$ 
by reading a child pointer of some Node $p$.
The update operation's arrival point at $\ell$ is the last configuration $C$ such that
(1) $C$ precedes the update's read of $\ell$ from a child pointer of $p$
and (2) $p$ is on the search path for $k$ in $C$.
($C$ exists by \Cref{was-on-path}.)
%\here{technically, to match the intuition, we should pick $C$ to be the earliest time during the operation when $\ell$ is on the search path for $k$}
\end{enumerate}

We define the remaining arrival points inductively.
Assuming the arrival points are defined for a prefix of the execution $\alpha$,
we describe the arrival points associated with the next step $s$ of the execution.

We first consider steps that change the Node tree.
\begin{enumerate}
\setcounter{enumi}{1}
\item\label[part]{a-succ-ins}
Consider a step $s$ that changes the Node tree to perform an
\op{Insert}$(k)$.  Then $s$ replaces a leaf $\ell$ of the Node tree by
an internal Node \x{new} with two leaf children, \x{newLeaf} which contains $k$, and $\ell'$, which contains the same key as $\ell$ (see \Cref{succ-ins-fig}).
The step $s$ is the arrival point at \x{new} of
all operations whose arrival points at $\ell$ precede $s$ (in the order they arrived at $\ell$).
Step $s$ is also the arrival point for each of these operations at one of \x{newLeaf} or $\ell'$ (in the same order):
those operations on keys less than \x{new}.\x{key} go to the left child of \x{new},
and the rest go to the right child.
Finally, the last arrival point placed at $s$ is for \op{Insert}$(k)$ at both \x{newLeaf} and \x{new}.
\item\label[part]{a-succ-del}
Consider a step $s$ that changes the Node tree
to perform a \op{Delete}$(k)$.  Then $s$ removes a leaf $\ell$ containing $k$ and
$\ell$'s parent $p$ from the Node tree (see \Cref{succ-del-fig}) by changing the appropriate child pointer of $\ell$'s grandparent $gp$ from $p$ to $\ell$'s sibling $sib$.
The step $s$ is the arrival point of the \op{Delete}$(k)$ at $sib$ and at every descendant of $sib$
on the search path for $k$.
Furthermore,
for each update operation that has an arrival point at $l$,
$s$ is the arrival point of the operation at $sib$ and at every descendant of $sib$ that is on the search path for the key of the operation.
If $s$ is the arrival point of multiple operations at a Node,
they are ordered as they were ordered at $\ell$, with the \op{Delete}($k$) last.
\end{enumerate}
Finally, we consider a successful \CAS\ $s$ performed by a \op{Refresh} $R$.
\begin{enumerate}
\setcounter{enumi}{3}
\item\label[part]{a-left}
Consider a \op{Refresh}($x$) $R$ that performs a successful \CAS\ $s$ on $x.\x{version}$ 
at line \ref{cas-refresh'}.
Let $x_L$ be the left child of $x$ read by $R$'s last execution of line \ref{read-left-child}.
Step $s$ is the arrival point at $x$ of all operations
that have an arrival point at $x_L$ prior to $R$'s last read at line \ref{read-left-version} and do not already have an arrival point at $x$ prior to $s$.
\item\label[part]{a-right}
Consider a \op{Refresh}($x$) $R$ that performs a successful \CAS\ $s$ on $x.\x{version}$
at line \ref{cas-refresh'}.
Let $x_R$ be the right child of $x$ read by $R$'s last execution of line \ref{read-right-child}.
Step $s$ is the arrival point at $x$ of all operations
that have an arrival point at $x_R$ prior to $R$'s last read at line \ref{read-right-version} and do not already have an arrival point at $x$ prior to $s$.
\end{enumerate}
If multiple operations' arrival points at an internal Node $x$ are at the same successful \CAS\ on $x.\x{version}$, we order them as follows:  
first the operations described in \cref{a-left} in the order of their arrival points at $x_L$ and 
then the operations described in \cref{a-right} in the order of their arrival points at $x_R$.

For operations whose arrival points are defined by \Cref{a-fail}, the associated response is \op{false}.
The \op{Insert}($k$) whose arrival point is defined by \Cref{a-succ-ins} and the \op{Delete}($k$) whose arrival point is defined in \Cref{a-succ-del} get an associated response of \true.
Whenever arrival points are copied from one Node to another by any of Parts \ref{a-succ-ins} to \ref{a-right}, the associated responses are copied as well.
\end{definition}

As in \Cref{static-proof}, the arrival points define a sequence of operations $Ops$ at each Node.
Now, we also attach a boolean response to each of the operations in the $Ops$ sequence
as we define the arrival points.
%\here{Would it be better to talk about a sequential history (of invocation-response pairs) rather than a sequence of operations?  PRobably, but fix it later.}

\begin{definition}
For each configuration $C$ and Node $x$, let
$Ops(C,x)$ be the sequence of update operations with arrival points at $x$ 
that are at or before $C$, in the order of their arrival points at~$x$.
Each operation in the sequence is annotated with a boolean response.
Let $Ops^*(C,x)$ be the sequence of update operations with arrival points at $x$ that are strictly before $C$, in the order of their arrival points at $x$.
Let $Ops(C,x,k)$ be the subsequence of $Ops(C,x)$ consisting of operations with argument~$k$.
\end{definition}

Since the \op{Refresh} routine works in the same way as in \Cref{static}, the following
result has an identical proof to \Cref{non-overlapping}.

\begin{observation}\label{non-overlapping-bst}
If two calls to \op{Refresh} on the same Node both perform successful \op{CAS} steps,
then one performs the read on line \ref{read-old'} after the \op{CAS} on line \ref{cas-refresh'} of the other.
\end{observation}

%We wish to prove that the linearization point of each completed update operation exists
%and that the linearization point of any update is within its execution interval.
The following claims are used to prove that no update is ``lost'' by the 
propagation technique if a Node where it has arrived is deleted from the tree.

\begin{invariant}\label{downward-closed}
For any configuration $C$ and any internal Node $x$,
each operation in $Ops(C,x)$ is also in $Ops(C,y)$ for some child $y$ of $x$ in $C$.
Moreover, the set of operations in the $Ops$ sequences of the children of $x$ can only grow.
\end{invariant}
\begin{proof}
The invariant holds vacuously in the initial configuration because no operations have arrival points.
Assume that the claims hold up to some configuration $C$.  We show that they hold up to the next configuration $C'$.

\Cref{a-fail} of \Cref{arrival-bst} adds update operations to the $Ops$ sequence only of leaf Nodes, so it trivially preserves the claims.

\Cref{a-succ-ins} of \Cref{arrival-bst} ensures that when an insertion occurs,
all operations added to the new internal Node's $Ops$ sequence
are also added to the $Ops$ sequence of one of its children (and that child is a leaf).  
(See \Cref{succ-ins-fig}.)
Moreover, no violation of the invariant
is created at the Node whose child pointer changes, since all arrival points of the replaced
leaf are transferred to the new internal Node that replaces it.

\Cref{a-succ-del} of \Cref{arrival-bst} ensures that when a deletion occurs,
all operations added to any internal Node's $Ops$ sequence
are also added to one of its children's $Ops$ sequence.
(See \Cref{succ-del-fig}.)
Moreover, no violation of the invariant
is created at the Node whose child pointer changes, since all arrival points of the old
child came from one of its children (by the induction hypothesis) and all of these are
transferred to the new child. 

Consider the arrival points added to a Node $x$ by
\Cref{a-left,a-right} of \Cref{arrival-bst} when the \CAS\ of a \op{Refresh} operation updates $x.\x{version}$.
All of the newly added operations had arrival points at $x$'s children when their \x{version}
fields were previously read by the \op{Refresh}, by the induction hypothesis.
By the second claim of the induction hypothesis, 
these operations still have arrival points
at one of $x$'s children at or before $C$ (even if $x$'s children have changed), 
so the first claim is satisfied in~$C'$.
\end{proof}

Next, we show that a successful \op{Refresh} propagates operations up the Node tree.

\begin{lemma}\label{arrive-refresh}
Suppose Node $y$ is a child of Node $x$ at a configuration $C$ and that an update operation
$op$ has an arrival point at $y$ at or before $C$.
If \op{Refresh}($x$) reads $x.\x{version}$ at line \ref{read-old'} after $C$ and performs a successful 
\CAS\ on line \ref{cas-refresh'}  then
$op$ has an arrival point at $x$ at or before the \CAS.
\end{lemma}
\begin{proof}
Assume $y$ is the left child of $x$; the case where it is the right child is symmetric.
Let $y'$ be the left child of $x$ that the \op{Refresh} reads at line \ref{read-left-child}.
By the second claim of \Cref{downward-closed}, $op$ has an arrival point at $y'$ before this read.
By \Cref{a-left} of \Cref{arrival-bst}, $op$ has an arrival point at $x$
at or before the successful \CAS\ of the \op{Refresh}.
\end{proof}

Now, we show that a double \op{Refresh} propagates operations, whether one of them performs a successful \CAS\ or not.
The following proof is similar to the induction step in the proof of \Cref{before-end}.

\begin{lemma}\label{arrive-double-refresh}
Suppose Node $y$ is a child of Node $x$ at a configuration $C$ and that an update operation
$op$ has an arrival point at $y$ before $C$.
If a process executes the double \op{Refresh} at lines \ref{refresh-1'}--\ref{refresh-2'} on $x$ after $C$ then
$op$ has an arrival point at $x$ at or before the end of the double \op{Refresh}.
\end{lemma}
\begin{proof}
If either \op{Refresh} has a successful \CAS, the claim follows from \Cref{arrive-refresh}.
If the \CAS\ steps performed by both calls $R_1$ and $R_2$ to \op{Refresh}($x)$ fail,
then there must have been two successful \CAS\ steps $c_1$ and $c_2$ on $x.\x{version}$
during $R_1$ and $R_2$, respectively.  (See \Cref{before-end-fig}.)
By \Cref{non-overlapping-bst}, the invocation $R$ of \op{Refresh}($x$) that performed $c_2$
must have read $x.\x{version}$ at line \ref{read-old'} \emph{after} $c_1$, which is after
the beginning of $R_1$.
Thus, by \Cref{arrive-refresh} applied to $R$, $op$ has an arrival point at $x$
no later than $c_2$, which is before the end of $R_2$.
\end{proof}

In \Cref{arrival-base,arrived-first-new,ancestor-subset,before-end-bst}, we consider an update operation $op$ on a key $k$ and show that it has an arrival point at the root before it terminates.
Let $x_1,\ldots,x_m$ be the Nodes on the local stack (from
the newest pushed to the oldest) when $op$ calls \op{Propagate}.
The first arrival point of $op$ at any Node is defined by \Cref{a-fail}, \ref{a-succ-ins}
or \ref{a-succ-del} of \Cref{arrival-bst}.
Let $C_0$ be the first configuration at or after this arrival point.
Since $x_1$ is on the stack, it is an internal Node.  Let $x_0$ be the left child of $x_1$ in $C_0$ if
$k<x_1.\x{key}$ or the right child of $x_1$ otherwise.
In \Cref{before-end-bst}, we wish to show  by induction on $i$ that before $op$'s \op{Propagate} completes its double \op{Refresh}
on $x_i$, $op$ has an arrival point at $x_i$.
We first prove the base case in \Cref{arrival-base} by showing
that $op$ has an arrival point at $x_0$ before \op{Propagate} is called.
\Cref{arrived-first-new,ancestor-subset} are useful for the induction step of \Cref{before-end-bst}, which 
shows that if a double \op{Refresh} is called after $op$ has an arrival point
at $x_{i-1}$, then before the double \op{Refresh} completes, $op$ will have an arrival point at~$x_i$.

\begin{lemma}
\label{arrival-base}
\Cref{a-fail}, \ref{a-succ-ins}
or \ref{a-succ-del} of \Cref{arrival-bst} defines the arrival point of $op$ at $x_0$ at or before $C_0$ and $C_0$ is before $op$'s call to \op{Propagate}.  Moreover, $x_0$ is on the search path for $k$ in $C_0$.
\end{lemma}
\begin{proof}
We consider several cases.

First, suppose $op$ is an \op{Insert}($k$) whose test on line \ref{ins-test} returns \true\ because it found a leaf $\ell$ that contains $k$.
\Cref{a-fail} of \Cref{arrival-bst} defines the arrival point of $op$ at $\ell$
to be the latest configuration before the \op{Insert} reads $\ell$ as the child of $x_1$
in which $x_1$ is on the search path for $k$.
By definition, $C_0$ is that configuration.
Either $C_0$ is immediately before the read of $x_0$ from one of $x_1$'s children
or $C_0$ is an earlier configuration immediately before the step that removes $x_1$ from the BST. 
%\here{does this require more justification?}
In the latter case, $x_1$ is marked in $C_0$, so its child pointers
cannot change thereafter (by \Cref{change-unmarked}),
so $x_0$ is already the child of $x_1$ in the search path for $k$ in $C_0$.
Either way, $\ell$ is the next Node after $x_1$ on the search path for $k$ in $C_0$,
so $x_0=\ell$, and $C_0$ is the arrival point of $op$ at $x_0$.
Finally, $C_0$ is during $op$'s search, by \Cref{was-on-path}, so it is before $op$ calls \op{Propagate} on line \ref{ins-prop-1}.

The claim can be proved similarly if $op$ is a \op{Delete}($k$) whose
test on line \ref{del-test} returns \true\ because it found a leaf $\ell$ that does not contain $k$:  in this case, $x_0$ is again $\ell$.

Now suppose $op$ is an \op{Insert}($k$) 
that successfully adds a leaf containing $k$ to the Node tree.
Then, $C_0$ is the configuration immediately after the successful \CAS\
that adds $k$ to the BST.
This step changes a child pointer of $x_1$, the top Node on $op$'s stack,
to add three new Nodes to the Node tree.
So, $x_0$ is the new internal
Node created on line \ref{new-triangle} of $op$.
The successful \CAS\ is the 
arrival point of the \op{Insert}
at $x_0$, by \Cref{a-succ-ins} of \Cref{arrival-bst}, and $C_0$ is the configuration after this \CAS.
This \CAS\ may be performed by $op$ itself or a helper, but, either way, it must be
done before the \op{Insert}($k$) reaches line \ref{ins-prop-2}, since it is the linearization
point of the \op{Insert} in the unaugmented BST.
It remains to show that $x_0$ is on the search path for $k$ in $C_0$.
By \Cref{was-on-path}, $x_1$ was on the search path for $k$ in some earlier configuration.
By \Cref{change-unmarked}, $x_1$ is still reachable when the \CAS\ occurs.
By \Cref{still-on-path}, this means that $x_1$ is still on the search path for $k$ when the CAS occurs,
so $x_0$ is on the search path for $k$ in the configuration $C_0$ after this \CAS.

Finally, suppose $op$ is a \op{Delete}($k$) 
that successfully removes a leaf containing $k$ from the Node tree.
Then, $C_0$ is the configuration immediately after the successful \CAS\
that deletes $k$ from the BST.
This step changes a child pointer of $x_1$, the top Node on $op$'s stack,
to the sibling of the deleted leaf.
So, $x_0$ is this sibling.
The \CAS\ is the arrival point of the \op{Delete} at $x_0$, by
\Cref{a-succ-del} of \Cref{arrival-bst}.
For the same reason as for \op{Inserts}, this \CAS\ is prior to $op$'s call to \op{Propagate} on line \ref{del-prop-2}.
The argument that $x_0$ is on the search path for $k$ in $C_0$ is also similar:  $x_1$ is on the
search path for $k$ in $C_0$, so $x_0$ is too.
\end{proof}

Since Nodes $x_1,\ldots,x_m$ were on the stack, they are all internal Nodes. This ensures
that the Node $x$ in the following claim is well-defined.

\begin{lemma}\label{arrived-first-new}
Consider any configuration $C$ at or after $C_0$.
Let $x$ be the first node in the search path for $k$ in $C$ that is not in $\{x_1,\ldots,x_m\}$.
Then,
$op$ is in $Ops(C,x)$.
\end{lemma}
\begin{proof}
The claim is true in $C_0$, since \Cref{arrival-base} ensures $x_0$ is on the search path for $k$ in $C_0$ and  $op$ is in $Ops(C_0,x_0)$.

As an induction hypothesis, assume the claim holds in some configuration $C$ to prove that it holds in the next configuration $C'$.
Let $P$ be the search path for $k$ in $C$ and $P'$ be the search path in $C'$.
Let $x$ and $x'$ be the first Nodes on $P$ and $P'$, respectively, that are not in $\{x_1,\ldots,x_m\}$.
If $x=x'$, then
the claim trivially follows from the induction hypothesis.
So, assume for the rest of the proof that $x\neq x'$.
The step from $C$ to $C'$ must have changed the child pointer
of some Node $x_i$ to~$x'$.

If this step changes the Node tree to perform an insertion, 
$x_i$'s child before the \CAS\ was a leaf and therefore is not in the
set $\{x_1,\ldots,x_m\}$ of internal Nodes.
Thus, that leaf must have been $x$, the first Node on $P$ 
that is not in $\{x_1,\ldots,x_m\}$.
By the induction hypothesis, $op$ is in $Ops(C,x)$.
By the second claim of \Cref{downward-closed}, $op$ is in $Ops(C',x')$.

If the step changes the Node tree to perform a deletion, the step 
removes a leaf Node $\ell$ and $\ell$'s parent $p$ from the Node tree by
changing a child pointer of $x_i$ from $p$ to $\ell$'s sibling $x'$.
So, prior to the \CAS, $x$
must have been $p$ (if $p\notin \{x_1,\ldots,x_m\}$) or $\ell$ (otherwise).
We first argue that $op$ is in either $Ops(C,x')$ or $Ops(C,\ell)$.
If $x=p$, then $op$ is in $Ops(C,p)$ by the induction hypothesis, and in one of 
$Ops(C,\ell)$ or $Ops(C,x')$ by \Cref{downward-closed}.
If $x=\ell$, then $op$ is in $Ops(C,\ell)$ by the induction hypothesis.
Finally, we argue that $op$ is in $Ops(C',x')$.
This is trivial if $op$ is in $Ops(C,x')$.
If $op$ is in $Ops(C,\ell)$ then \Cref{a-succ-del} of \Cref{arrival-bst}
ensures that it is in $Ops(C',x')$.
\end{proof}

%\here{reverse order of $x_1...x_m$; some of this discussion can be moved to discussion of original BST; x1 is the root xm is the parent of the leaf where the search arrived}
Next, we consider the nodes $x_1,\ldots,x_m$ on the stack on which
$op$'s \op{Propagate} will perform a double \op{Refresh}.
In the proof of \wasOnPath\ in \cite{EFHR14}, it is shown that there is a sequence of configurations
$D_{m-1}, \ldots, D_1$ where $D_i$ is at or before $D_{i-1}$ for all $i$, such that
$x_i$ and $x_{i-1}$ are consecutive Nodes on the search path for $k$ in $D_{i-1}$.
Intuitively, these configurations can be defined inductively as follows.
$D_{m-1}$ is when $x_{m-1}$ is read from a child of the permanent 
root node $x_m$.
Once $D_i$ is defined, we consider two cases:
if $x_i$ is still on the search path for $k$ when its child pointer to $x_{i-1}$
is read during the search, then let $D_{i-1}$ be this read; otherwise,
let $D_{i-1}$ be the configuration before $x_i$ was deleted from the search path.
In the latter case, $x_i$'s child already pointed to $x_{i-1}$ at $D_{i-1}$ because
$x_i$'s child pointers cannot be changed after it is deleted, by \Cref{change-unmarked}.

\begin{lemma}\label{ancestor-subset}
In any configuration at or after $D_i$,
if $x_i$ is still in the search path for $k$, then $x_i$'s ancestors
are a subset of $x_{i+1},\ldots,x_m$.
\end{lemma}
\begin{proof}
The claim holds vacuously for the root Node $x_m$.
Assume the claim holds for $x_{i+1}$ to prove it for $x_i$.
In the configuration $D_i$, $x_{i+1}$ and $x_i$ are consecutive Nodes
on the search path for $k$.
Since this configuration is after $D_{i+1}$, the ancestors of $x_{i+1}$
are a subset of $\{x_{i+2},\ldots,x_m\}$.
Thus the claim holds for $x_i$ in $D_i$.
Any step that modifies the Node tree after $D_i$ preserves the claim
because it can only splice out a Node (for a deletion) or replace a leaf.
(See Figure \ref{tree-changes}.)
%\here{may want to cite proofs from \cite{EFHR14} for some of this}
\end{proof}

The next lemma implies that each completed update operation
has a linearization point, which is before the update terminates.

\begin{lemma}
\label{before-end-bst}
Suppose an update operation $op$ calls \op{Propagate}, which does a double \op{Refresh} on
Nodes $x_1, x_2, \ldots, x_m$.  Then $op$ has an arrival point at
$x_i$ before the end of the $i$th iteration of the loop in \op{Propagate}.
\end{lemma}
\begin{proof}
We prove by induction on $i$ that for $0\leq i\leq m$, $op$ has an arrival point at $x_i$
before the end of $i$ iterations of the loop in \op{Propagate}.
When $i=0$, this follows from \Cref{arrival-base}.

For the induction step, assume the claim holds for $x_0,\ldots,x_{i-1}$ and
consider the $i$th iteration of the loop in \op{Propagate}.
If one of $x_0,\ldots,x_{i-1}$ is a child of $x_i$
at the beginning of the $i$th iteration of the loop, then the claim follows from 
\Cref{arrive-double-refresh}.
Otherwise, by \Cref{ancestor-subset}, the child of $x_i$ at the beginning of the $i$th iteration of the loop
is the first Node on the search path for $k$ that is not in $\{x_1,\ldots,x_m\}$.
By \Cref{arrived-first-new}, $op$ has arrived at that Node before the $i$th iteration of the loop,
so the claim follows from \Cref{arrive-double-refresh}.
\end{proof}

It follows from \Cref{before-end-bst} that $op$ has an arrival point at the root Node
before it terminates.  Now it remains to show that the arrival point at the root Node
is after $op$ is invoked, which is quite easy.

\begin{lemma}\label{after-start-bst}
The arrival point of an operation at any Node is after the operation is invoked.
\end{lemma}
\begin{proof}
The first arrival point of an \op{Insert} at any Node is defined either
by \Cref{a-fail} or \Cref{a-succ-ins} of \Cref{arrival-bst}.
(All other parts define arrival points of \op{Inserts} that have already arrived at other Nodes.)
For \Cref{a-succ-ins}, the arrival point is when the tree is updated to perform
the \op{Insert}.  
In \cite{EFHR14}, this modification to the tree is used as the linearization point of the \op{Insert},
so it was shown in \cite[\realTime]{EFHR14} that it is during the \op{Insert}.
For \Cref{a-fail}, the arrival point is after the start of the \op{Insert}, by definition.
The claim follows for \op{Insert} operations.

The proof for \op{Delete} operations is similar.
\end{proof}

Recall that the linearization point of an update operation is its arrival point at the root Node.
It follows from \Cref{before-end-bst,after-start-bst} that the linearization 
point of each update operation is within the execution interval of that operation.
%\here{perhaps explain a little; also we should probably talk about the fact that an operation cannot have two different arrival points at a node, but that requires stuff later; so maybe here define lin point as the first arrival point at the root}

\subsection{Linearization is Consistent with Responses}
\label{bst-match-proof}

In a configuration $C$, we say a Node is \emph{reachable} if there is a path of child
pointers from the root Node to it.

\begin{invariant}\label{left-right}
In every configuration $C$, for every internal Node $x$ that is reachable in $C$,
\begin{itemize}
\item
for every Node $x_L$ that was in $x$'s left subtree at or before $C$,
every operation in $Ops(C,x_L)$ has a key less than $x.\x{key}$, and
\item
for every Node $x_R$ that was in $x$'s right subtree at or before $C$,
every operation in $Ops(C, x_R)$ has a key greater than or equal to $x.\x{key}$.
\end{itemize}
\end{invariant}
\begin{proof}
Initially, the invariant holds vacuously, since all $Ops$ sequences are empty.

Assume the invariant holds in all configurations up to $C$ and let $s$ be 
a step from configuration $C$ to another configuration $C'$.
Consider any internal Node $x$ that is reachable in $C'$.
To prove the claim holds for $x$ in $C'$, we 
need only consider a step $s$ if it modifies the subtree of the Node tree rooted at $x$,
or if it modifies an $Ops$ sequence of a Node in that subtree.
We prove the first claim for the left subtree of $x$; the arguments for the right subtree are symmetric in all cases.

First, consider a successful \CAS\ step $s$ that modifies the 
left subtree of $x$ to perform an \op{Insert}($k$). 
This step $s$
changes a child pointer of a reachable Node $p$ from a leaf $\ell$ (where $\ell$ is in the left subtree of $x$) to a new Node
\x{new} that has two leaf children $\ell'$ and \x{newLeaf}.  (See \Cref{succ-ins-fig}.)
By \Cref{a-succ-ins} of \Cref{arrival-bst},
the step $s$ is the arrival point at both \x{new} and the appropriate child of \x{new} of
all operations whose arrival points at $\ell$ precede $s$.
By the induction hypothesis, since $\ell$ was in the left subtree of $x$ in $C$,
all of these operations had keys less than $x.\x{key}$.
Step $s$ is also the arrival point at \x{newLeaf} and \x{new} of the \op{Insert}
of $\x{newLeaf}.\x{key}=k$.
Since the Node tree is always a BST by \Cref{basic-bst}, it follows that $k<x.\x{key}$.

Next, consider a successful \CAS\ step $s$ that modifies the left subtree of $x$ 
to perform a \op{Delete}($k$).  
This step removes a leaf $\ell$ and its parent $p$ (which is in the left subtree of $x$) 
by changing a child pointer of $p$'s parent $gp$, which is reachable, from $p$ to $\ell$'s sibling $sib$.  
(See \Cref{succ-del-fig}.)
By \Cref{a-succ-del} of \Cref{arrival-bst}, $s$ is the arrival point at $sib$ and some of its descendants
of operations in $Ops(C,\ell)$.
Since $\ell$ is in the left subtree of $x$ in $C$, the induction
hypothesis ensures that all of these operations have keys less than $x.\x{key}$.
Step $s$ is also the arrival point of the \op{Delete}$(k)$ at $sib$ and some of its descendants.
Since the Node tree is always a BST (by \Cref{basic-bst}) and the key $k$ appeared in Node $\ell$ in the left subtree of $x$,
it follows that $k < x.\x{key}$.

Next, consider a successful \CAS\ step $s$ on the \x{version} field of some Node
$y$ that was in $x$'s left subtree at some time before $C$.
This step is performed by a \op{Refresh}($y$).  
Let $y_L$ and $y_R$ be the Nodes that the \op{Refresh} reads 
from $y$'s child fields at lines \ref{read-left-child} and \ref{read-right-child}, respectively.
If $y$ is still reachable when $y_L$ is read from its left child field, then 
$y$ is in the left subtree of $x$ at that time, since the only types of changes
to the Node tree that can happen are those shown in \Cref{tree-changes}.
%\here{Previous sentence probably needs a better explanation}
So, $y_L$ is also in the left subtree of $x$ at that time.
If $y$ is not reachable when $y_L$ is read from its left child field,
then $y_L$ was the left child of $y$ when $y$ was deleted (since $y$'s child
fields cannot change after it is deleted, by \Cref{change-unmarked}).
So, $y_L$ was in the left subtree of $x$ just before $y$ was deleted.
Either way, $y_L$ was in the left subtree of $x$ before $C$.
The argument for $y_R$ is symmetric, so we can apply the induction hypothesis to both $y_L$ and $y_R$.
Step $s$ is the arrival point at $y$ of operations
that have an arrival point at $y_L$ or $y_R$ prior to the reads 
at line \ref{read-left-version} or \ref{read-right-version}, respectively.
By the induction hypothesis, all such operations have keys less than $k$.

Finally we consider arrival points that are at configuration $C'$ itself (and not at step $s$).
By \Cref{a-fail} of \Cref{arrival-bst}, $C'$ can be the arrival point
of an \op{Insert}($k$) operation at a leaf $\ell$ 
if its search arrived at $\ell$ containing $k$, and $\ell$ is reachable in $C'$.
If $\ell$ was in the left subtree of $x$ before $C'$,
then it still is in the left subtree of $x$ because $x$ and $\ell$ are reachable in $C'$.
By the BST property of the Node tree (\Cref{basic-bst}), $k<x.\x{key}$.
Similarly, $C'$ can be the arrival point at a leaf $\ell$ of a \op{Delete}($k$)
if its search arrived at a leaf $\ell$ that does not contain $k$, and $\ell$ is reachable in $C'$.
If $\ell$ was in the left subtree of $x$ before $C'$,
then it still is in the left subtree of $x$ because $x$ and $\ell$ are reachable in $C'$.
%\here{need to explain this--same argument as in previous parag so maybe extract a lemma}
So, by the BST property of the Node tree, $\ell.\x{key} < x.\x{key}$.
\end{proof}

Leaf nodes satisfy the following claim because it holds whenever a leaf Node is
created (on line \ref{new-node'}) and the fields of leaf Nodes and Version objects are never changed.
\begin{observation}\label{leaf-version}
For each leaf Node $x$, $x.\x{version}$ always points to a Version with key $x.\x{key}$ and no children.
\end{observation}

\begin{invariant}\label{ops-prefix}
For all configurations $C$ and all Nodes $x$ that are reachable in $C$, if $x_L$ and $x_R$ are the left and right child of $x$ in $C$, then
\begin{itemize}
\item for keys $k<x.\x{key}$, $Ops(C,x,k)$ is a prefix of $Ops(C,x_L,k)$, and
\item for keys $k\geq x.\x{key}$, $Ops(C,x,k)$ is a prefix of $Ops(C,x_R,k)$.
\end{itemize}
\end{invariant}

\begin{proof}
Initially, the invariant holds vacuously, since all $Ops$ sequences are empty.
We must show that the invariant is preserved 
by any change to $Ops$ sequences of a Nodes,
and by any change to the structure of the Node tree.

We first consider changes to the $Ops$ sequences of Nodes.
Adding an arrival point to a leaf Node clearly preserves the invariant.
\Cref{a-fail} of \Cref{arrival-bst} only adds arrival points to leaf Nodes.
Whenever \cref{a-succ-ins} or \cref{a-succ-del} of \Cref{arrival-bst}
adds an arrival point of an operation
to an internal Node, it also adds adds an arrival point of the operation to the appropriate child of that Node, so the invariant is preserved.
It remains to check that a successful \CAS\ step of a \op{Refresh}($x$) that updates $x.\x{version}$ preserves the invariant at $x$.
 \Cref{a-left,a-right} of \Cref{arrival-bst} may append operations to $Ops(C,x,k)$.
By \Cref{left-right}, the new operations can only come from $x_L$ if $k<x.\x{key}$ or $x_R$ if $k\geq x.\x{key}$.
Thus, $Ops(C,x,k)$ simply becomes a longer prefix of $Ops(C,x_L,k)$ or $Ops(C,x_R,k)$.

Next, we show that changes to the structure of the Node tree preserve the claim.
Consider a successful \CAS\ step that modifies the 
Node tree to perform an \op{Insert} operation. 
It changes a child pointer of a reachable Node $p$ from a leaf $\ell$ to a new Node
\x{new}.  (See \Cref{succ-ins-fig}.)
Since the $Ops$ sequence of $\ell$ is transferred to \x{new} by \Cref{a-succ-ins} of \Cref{arrival-bst},
this change preserves the claim.

Now consider a successful \CAS\ step that modifies the Node tree to perform a
\op{Delete} operation.
Let $C$ be the configuration before the step and $C'$ be the configuration after.
This step removes a leaf $\ell$ and its parent $p$ 
by changing a child pointer of $p$'s parent $gp$, which is reachable, from $p$ to $\ell$'s sibling $sib$.
Assuming the claim holds in $C$, we must show that $Ops(C',gp,k)$ is a prefix of $Ops(C',sib,k)$ for all $k$.  
Without loss of generality, assume $\ell$ is $p$'s left child and $p$ is $gp$'s right child, as shown in \Cref{succ-del-fig}.  (The other cases are argued in exactly the same way.)

First, consider keys $k\geq p.\x{key}\geq gp.\x{key}$.
Since the claim holds in~$C$, we have
$Ops(C,gp,k)$ is a prefix of $Ops(C,p,k)$, which is a prefix of $Ops(C,sib,k)$.
Thus, $Ops(C',gp,k)=Ops(C,gp,k)$ is a prefix of $Ops(C,sib,k)=Ops(C',sib,k)$ as required.

Now, consider keys $k$ satisfying $gp.\x{key}\leq k<p.\x{key}$.
Since the claim holds in $C$, 
$Ops(C,gp,k)$ is a prefix of $Ops(C,p,k)$, which is a prefix of $Ops(C,\ell,k)$.
So, $Ops(C,gp,k)$ is a prefix of $Ops(C,\ell,k)$.
Moreover, $Ops(C,sib,k)$ is empty, by \Cref{left-right}.
\Cref{a-succ-del} of \Cref{arrival-bst} ensures that the step
transfers the arrival points at $\ell$ to $sib$, so we have $Ops(C',sib,k) = Ops(C,\ell,k)$.
So, we have $Ops(C',gp,k)=Ops(C,gp,k)$ is a prefix of $Ops(C,\ell,k)=Ops(C',sib,k)$, as required.
\end{proof}

The following crucial invariant shows that in every configuration $C$,
the Version tree stored in each Node $x$ accurately reflects
all the operations that have arrival points at $x$ at or before $C$: 
the set of keys that would be obtained by performing the operations in $Ops(C,x)$ sequentially
is exactly the set of keys that are actually
stored in the leaves of the Version tree that $x.\x{version}$ points to in configuration $C$.
When we talk about the set of keys in the leaves,
we include only keys from $U$; we exclude the dummy leaves containing $\infty_1$ and $\infty_2$.

\begin{invariant}\label{accurate}
In every configuration $C$, for every Node $x$ that is reachable in $C$,
\begin{enumerate}[label=(\arabic*)]
\item\label{abstract}
the Version tree rooted at $x.\x{version}$ is a BST whose leaves contain exactly
the keys
that would be in a set after performing $Ops(C,x)$ sequentially, and
\item\label{responses}
the responses recorded in $Ops(C,x)$ are consistent with executing the operations sequentially.
\end{enumerate}
\end{invariant}
\begin{proof}
In the initial configuration, the Version tree of each Node contains no keys (see Figure \ref{bst-init}), 
and all $Ops$ sequences are empty, so the claims hold vacuously.

Assume the claims hold for the prefix of the execution up to some configuration $C$.
We consider the next step $s$ after $C$, and show the claims hold in the resulting
configuration $C'$.
We prove this in two parts:  first, we show that 
a Node $x$'s Version tree in $C'$ contains the keys that should be present after doing
the sequence of updates $Ops^*(C',x)$ (i.e., that the Version tree reflects the operations whose arrival points at $x$ are at step $s$), and then show that all operations
whose arrival points at $x$ are at the configuration $C'$ have no effect on the set of keys that should
be stored in the Version tree rooted at $x$ (i.e., that they are updates that should return false).
We need only consider a step $s$ 
that changes the data structure or adds an arrival point to some Node $x$ (so that $Ops(C',x)\neq Ops(C,x)$).
We consider several cases.

\begin{enumerate}[label={Case \arabic*}]
\item
Consider a successful \CAS\ step $s$ that performs an \op{Insert}($k$) operation.
This step
changes a child pointer of a reachable Node $p$ from a leaf $\ell$ to a new Node
\x{new} that has two leaf children $\ell'$ and \x{newLeaf}, which contains $k$.  (See \Cref{succ-ins-fig}.)
Let $k'$ be the key of $\ell$ and $\ell'$.
By \Cref{leaf-version}, $\ell$'s Version tree in $C$ contains a single key $k'$.
By induction hypothesis \ref{abstract}, performing $Ops(C,\ell)$ results in the set $\{k'\}$.
Hence, for any key $k''\neq k'$, $Ops(C,\ell,k'')$ does not end with an \op{Insert}.
We must check that the claim is satisfied at each of \x{new}, \x{newLeaf} and $\ell'$.
(For any other reachable Node, the claim is true by the induction hypothesis.)

By \Cref{a-succ-ins} of \Cref{arrival-bst}, 
$Ops^*(C',\x{new}) = Ops(C,\ell) \cdot \langle \op{Insert}(k):\true\rangle$.
So, performing this sequence of operations would yield the set with two elements $\{k,k'\}$
(since $k\neq k'$ because the \CAS\ would not be performed if the test on line \ref{ins-test} were true).
This set matches the keys in the Version tree rooted at $\x{new}.\x{version}$ in $C'$, 
and the \op{Insert}($k$) would indeed return \true\ in the sequential execution.

The operations of $Ops(C,\ell)$ are split among the $Ops^*$ sequences of the leaves \x{newLeaf} and $\ell'$ according
to whether their keys are less than \x{new}.\x{key} or not.
But since, for any $k''\neq k'$, $Ops(C,\ell,k'')$ does not end with an \op{Insert},
we see that performing $Ops^*(C',\ell')$ yields the set $\{k'\}$ and performing $Ops^*(C',\x{newLeaf})$
(which also includes an \op{Insert}($k$) at the end) yields the set $\{k\}$.
These sets match the keys in the Version trees rooted at $\ell'.\x{version}$ and \x{newLeaf}.\x{version},
respectively.
Claim \ref{responses} at $\ell'$ and \x{newLeaf} also follow immediately from induction hypothesis \ref{responses} applied to $\ell$.

\item\label{accurate-del}
Consider a successful \op{CAS} step $s$ that performs a \op{Delete}($k$) operation.  
This step removes a leaf $\ell$ containing key $k$ and its parent $p$  
by changing a child pointer of $p$'s parent $gp$, which is reachable, from $p$ to $\ell$'s sibling $sib$.  
(See \Cref{succ-del-fig}.)
Suppose $\ell$ is the left child of~$p$.  (The other case is symmetric.)
This step also adds arrival points of operations in $Ops(C,\ell)$ to Nodes in the subtree rooted at
$sib$, according to \Cref{a-succ-del} of \Cref{arrival-bst}.
We must show that adding these new operations to the $Ops$ sequences of Nodes in $sib$'s subtree
does not change the set of keys that should be stored in their Version trees.
In other words, we must show that for each such Node $z$, the set of keys after doing
$Ops(C,z)$ sequentially is the same as the set of keys after doing $Ops^*(C',z)$ sequentially.
We do this by arguing about each key $k'$ separately.

By \Cref{leaf-version}, $\ell.\x{version}$ is the root of a single-node Version tree containing only the key $k$.
Thus, by the induction hypothesis, performing the operations in $Ops(C,\ell)$ sequentially
yields the set $\{k\}$.
Hence, performing the operations in the sequence $\sigma=Ops(C,\ell)\cdot\langle \op{Delete}(k):\true\rangle$
yields an empty set and the \op{Delete}($k$) would return \true\ in this sequential execution.
Let $k'$ be any key that appears in the sequence $\sigma$.
Since performing the operations of $\sigma$ sequentially yields the empty set,
the last operation on $k'$ in $\sigma$ must be a \op{Delete}($k'$).

The step $s$ adds arrival points of all updates in $\sigma$ with key $k'$ to 
$sib$ and each of $sib$'s descendants along the search path for $k'$.
Let $z$ be one of those Nodes.
Since an operation with key $k'$ had an arrival point at $\ell$, which is the left child of $p$ in $C$,
we must have $k'< p.\x{key}$ by \Cref{left-right}.
Thus, since $z$ is in the right subtree of $p$ in $C$, no operations on $k'$ can appear in $Ops(C,z)$,
again by \Cref{left-right}.
By the induction hypothesis applied to $z$, this means that the Version tree rooted at $z.\x{version}$
does not contain $k'$.  Thus, claim \ref{abstract} is satisfied with respect to $k'$, since $Ops^*(C',z,k')$ ends with a \op{Delete}($k'$).
Moreover, by induction hypothesis \ref{responses}, the operations in $Ops(C,\ell,k')$ are annotated 
with the response they would get if they were executed sequentially.
Since there are no operations on $k'$ in $Ops(C,z)$, performing the operations of $Ops(C,\ell,k')$ after $Ops(C,z)$ would yield the same responses, so claim \ref{responses} is also satisfied with respect to key $k'$.

\item\label{accurate-refresh}
Consider a successful \op{CAS} step $s$ performed by line \ref{cas-refresh'} of an execution $R$ of \op{Refresh}($x$).
For any Node other than $x$, the claims are true by the induction hypothesis.
Let $x_L$ and $x_R$ be the children of $x$ read by $R$'s last execution of line \ref{read-left-child} and \ref{read-right-child}, respectively.
Let $C_L$ and $C_R$ be the configurations before $R$'s last reads of
$x_L.\x{version}$ and $x_R.\x{version}$ on line \ref{read-left-version} and \ref{read-right-version}.
Let $T_L$ and $T_R$ be the Version trees whose roots are read by those two reads.
Let $T'$ be the tree whose root is stored in $x.\x{version}$ by step $s$.

By the construction on line \ref{new-node'}, the root of $T'$ is a Version with key $x.\x{key}$ and
its left and right subtrees are $T_L$ and $T_R$.
By the induction hypothesis, $T_L$ is a BST whose leaves contain exactly the keys that would be in a set
after performing $Ops(C_L, x_L)$ sequentially.
Similarly, $T_R$ is a BST whose leaves contain exactly the keys that would be in a set after performing
$Ops(C_R,x_R)$ sequentially.
It follows from \Cref{left-right}, that all keys in $T_L$ are less than $x.\x{key}$ and all keys in $T_R$ are
greater than or equal to $x.\x{key}$.  Thus, $T'$ is a BST.

Let $k$ be any key.  To prove claim \ref{abstract}, we show that $k$ appears in a leaf of $T'$ if and only if
$Ops^*(C',x,k)$ ends with an \op{Insert}($k$).
Assume $k < x.\x{key}$.  (The case where $k\geq x.\x{key}$ has a symmetric argument.)
Then, $Ops^*(C',x,k) = Ops(C_L,x,k)\cdot \tau\cdot \phi$, where
$\tau$ is the sequence of operations on key $k$ whose
arrival points at $x$ are between $C_L$ and $C$, and $\phi$ is the sequence of operations on key $k$ whose arrival points are at the \CAS\ step $s$.
Since the \CAS\ on $x.\x{version}$ at line \ref{cas-refresh'} of the \op{Refresh} $R$ succeeds, 
$x.\x{version}$ does not change between $R$'s read of $x.\x{version}$ at line \ref{read-old'} and the \CAS.
In particular, it does not change between $C_L$ and $C$.
Thus, the arrival points of operations in $\tau$ can only be those defined by \Cref{a-succ-del} of \Cref{arrival-bst}.
It was argued in \ref{accurate-del}, above, that whenever the arrival point of an operation
is added to a Node
by \Cref{a-succ-del} of \Cref{arrival-bst}, the key of that operation does not already appear in
the Node's $Ops$ sequence, and moreover the last operation added to the Node's $Ops$ sequence
with that key is a \op{Delete}.
In other words, one of $Ops(C_L,x,k)$ or $\tau$ must be empty, and if $\tau$ is non-empty, then it ends
with a \op{Delete}($k$).
Similarly, if $\tau$ is non-empty, then $Ops(C_L,x_L,k)$ must be empty.

We consider two subcases.

%\here{need to update remainder of proof to deal with new claim: can we just use the prefix claim for one subcase?}

\begin{enumerate}[label={\ref{accurate-refresh}(\alph*)}]
\item
Consider the case where $\tau$ is empty.
Then, $Ops^*(C,x,k) = Ops(C_L,x,k) \cdot \phi$.
Since $Ops(C_L,x,k)$ is a prefix of $Ops(C_L,x_L,k)$ by \Cref{ops-prefix},
and the \CAS\ step $s$ adds arrival points to $x$ for any operations in $Ops(C_L,x_L,k)$ that 
are not in $Ops(C,x,k)$, we have $Ops^*(C,x,k)=Ops(C_L,x_L,k)$.
Applying the induction hypothesis \ref{abstract}
to $x_L$ at configuration $C_L$, the key $k$ appears in a leaf of the tree $T_L$ 
if and only if $Ops(C_L,x_L,k)$ ends with an \op{Insert}($k$).
Thus, $k$ appears in a leaf of the tree $T'$ if and only if $Ops^*(C,x,k)$ ends with an \op{Insert}($k$),
since the BST property of $T'$ ensures that $k$ can only appear in the left subtree $T_L$ of $T'$.
Moreover, since $Ops^*(C,x,k)=Ops(C_L,x_L,k)$, and the responses in $Ops(C_L,x_L,k)$ are
consistent with the sequential execution by induction hypothesis \ref{responses},
claim \ref{responses} holds for $x$ in $C$ also.
\item
Now, consider the case where $\tau$ is non-empty.
As mentioned above, $Ops(C_L,x_L,k)$ and $Ops(C_L,x,k)$ must then be empty.
Since $\phi$ consists of operations from $Ops(C_L,x_L,k)$, $\phi$ must also be empty.
Thus, $Ops^*(C',x,k) = Ops(C_L,x,k)\cdot \tau\cdot \phi = \tau$.
As mentioned above, $\tau$ must end with a \op{Delete}($k$).
Since $Ops(C_L,x_L,k)$ is empty,  induction hypothesis \ref{abstract} applied to $x_L$ in $C_L$
implies that $k$ does not appear in any leaf of $T_L$.
Thus, $k$ does not appear in any leaf of $T'$, since the BST property of $T'$ ensures that 
$k$ cannot appear in the right subtree $T_R$ of $T'$.
So, claim \ref{abstract} is satisfied with respect to $k$.
Since $\tau$ is obtained (via \Cref{a-fail} of \Cref{arrival-bst}) by copying
an $Ops$ sequence some Node $x'$ to $x$, claim \ref{responses} with respect to key $k$
follows from induction hypothesis \ref{responses} applied to that Node $x'$
when in the configuration before the copying is done.
\end{enumerate}
\end{enumerate}

This completes the proof that the leaves of the Version tree that $x.\x{version}$ points to in $C'$
are exactly the keys of the set obtained by doing the operations in $Ops^*(C',x)$.
It remains to prove that the operations whose arrival points are at $C'$ itself also preserve the invariant.
Only \Cref{a-fail} of \Cref{arrival-bst} assigns arrival points to configurations,
and it assigns arrival points only at leaf Nodes.
Thus, if $x$ is an internal Node, $Ops(C',x)=Ops^*(C',x)$ and the proof of the invariant is complete.
Suppose $x$ is a leaf Node.
By definition, all operations whose arrival points at a leaf $x$ whose key is $k$ are either
\op{Insert}($k$) operations or \op{Delete}($k'$) operations with $k'\neq k$, and the responses
recorded for these operations are all \false.
By \Cref{leaf-version}, $x.\x{version}$ points to a BST containing just one leaf with key $k$.
Since we have already shown that performing the operations of $Ops^*(C',x)$ sequentially
results in the set $\{k\}$, all the operations whose arrival points are at $C'$ would not affect
the set if they were performed sequentially after $Ops^*(C',x)$, and they would all return \false\ in this sequential execution.  
Thus, the set that results from performing $Ops(C',x)$ is also $\{k\}$.  This completes the proof.
\end{proof}

\begin{corollary}
Each update that terminates returns a response consistent with the linearization ordering.
\end{corollary}
\begin{proof}
When an \op{Insert} that returns \false\ at line \ref{ins-return-false}
or a \op{Delete} that returns \false\ at line \ref{del-return-false}
is first assigned an arrival point by \Cref{a-fail} of \Cref{arrival-bst},
the response value associated with it is \false

Similarly, when an \op{Insert} that returns \true\ at line \ref{ins-return-true}
or a \op{Delete} that returns \true\ at line \ref{del-return-true}
is first assigned an arrival point by \Cref{a-succ-ins} or \Cref{a-succ-del} of \Cref{arrival-bst},
the response value associated with it is \true.

In both cases, the response value is carried up along with the operation
to its arrival point at the root Node.
It follows from \Cref{accurate}.\ref{responses} that the response is consistent
with performing the operations sequentially in their linearization order.
\end{proof}

The following invariant is trivial, since every internal Version object satisfies the invariant when it 
is created (at line \ref{new-node'}) and its fields are immutable.
\begin{invariant}
For every Version $v$ that has children, $v.\x{sum} = v.\x{left}.\x{sum} + v.\x{right}.\x{sum}$.
\end{invariant}

The following corollary is a consequence of the fact that each leaf Version containing a valid key (created at line \ref{new-triangle}) has \x{sum} field 1 and each leaf Version containing a sentinel key (created at line \ref{bst-initialize} or \ref{new-triangle}) has \x{sum} field 0.
\begin{corollary}\label{sum-accurate}
For every Version $v$, $v.\x{sum}$ is the number of leaves in the tree rooted at $v$ that contain keys from $U$.
\end{corollary}

\begin{lemma}
The result returned by each query operation is consistent with the linearization.
\end{lemma}
\begin{proof}
If we linearize any query operation (including \op{Find}) when it reads a value $v$ from $\x{Root}.\x{version}$, it follows from
\Cref{accurate} that the tree of Version objects rooted at $v$ is a BST containing exactly the keys
that a set would contain if all operations linearized prior to the query are performed in their linearization order.
Moreover, by \Cref{sum-accurate}, the \x{sum} fields in all nodes of the tree rooted at $v$ are accurate for
guiding order-statistic queries.
Thus, the Version tree rooted at $v$ is an immutable order-statistic tree for the set of items present in the set at the linearization point of the query.  The claim follows.
\end{proof}

\bibliographystyle{plain}
\bibliography{augmented-tree.bib}

\end{document}